\documentclass[12pt]{article}
\usepackage{amsfonts}
\usepackage{amssymb}
\usepackage{amsthm}
\usepackage{amsmath}
\usepackage{bm}
\begin{document}

\renewcommand{\theequation}{\thesection.\arabic{equation}}

\title{The averaging of multi-dimensional Poisson brackets for
systems having pseudo-phases}

\author{A.Ya. Maltsev}

\date{
\centerline{\it L.D. Landau Institute for Theoretical Physics}
\centerline{\it 142432 Chernogolovka, pr. Ak. Semenova 1A,
maltsev@itp.ac.ru}}

\maketitle

{\hfill {\it Dedicated to the 75th birthday}}

{\hfill {\it of Professor S.P. Novikov}}

\vspace{0.5cm}

\begin{abstract}
We consider features of the Hamiltonian formulation of the
Whitham method in the presence of pseudo-phases. As we show,
an analog of the procedure of averaging of the Poisson bracket
with the reduced number of the first integrals can be suggested
in this case. The averaged bracket gives a Poisson structure
for the corresponding Whitham system having the form similar
to the structures arising in the presence of ordinary phases.
\end{abstract}

\section{Introduction. Hamiltonian structures in the Whitham method.}

 In this paper we consider the Hamiltonian formulation of the
Whitham method for multi-dimensional systems having some additional
property. Namely, we consider multi-dimensional systems which
possess the so-called ``pseudo-phases'' having special physical
or geometrical meaning. This property manifests itself in particular
in the definition of the multi-phase solutions of the corresponding
systems and in the form of the corresponding Whitham equations.
Our considerations here will be devoted to the Hamiltonian
formulation of the Whitham equations which will be connected
with the procedure of the averaging of multi-dimensional Poisson
structures in the presence of pseudo-phases.
So, we consider the evolutionary systems
\begin{equation}
\label{InSyst}
\varphi^{i}_{t} \,\, = \,\, F^{i} (\bm{\varphi}, \bm{\varphi}_{\bf x},
\bm{\varphi}_{\bf xx}, \dots ) \,\, \equiv \,\,
F^{i}(\bm{\varphi}, \bm{\varphi}_{x^{1}},
\dots, \bm{\varphi}_{x^{d}}, \dots )
\end{equation}
$i = 1, \dots, n$ ,
$\bm{\varphi} = (\varphi^{1}, \dots, \varphi^{n})$, with $d$ spatial
dimensions, and their $m$-phase solutions which are usually written in
the form
\begin{equation}
\label{mphasesol}
\varphi^{i} ({\bf x}, t) \,\, = \,\, \Phi^{i} \left(
{\bf k}_{1}({\bf U})\, x^{1} \, + \, \dots \, + \,
{\bf k}_{d}({\bf U})\, x^{d}
\, + \, \bm{\omega}({\bf U})\, t \, + \, \bm{\theta}_{0}, \,
{\bf U} \right)
\end{equation}
with some $2\pi$-periodic in each $\theta^{\alpha}$ functions
$$\Phi^{i} \left( \bm{\theta}, {\bf U} \right) \,\,\, \equiv \,\,\,
\Phi^{i} \left( \theta^{1}, \dots, \theta^{m}, \, {\bf U} \right) $$

 Here the functions
${\bf k}_{q} ({\bf U}) = (k^{1}_{q} ({\bf U}), \dots,
k^{m}_{q} ({\bf U}))$ 
and $\bm{\omega} ({\bf U}) = (\omega^{1} ({\bf U}), \dots,
\omega^{m} ({\bf U}))$ represent the ``wave numbers'' and
the ``frequencies'' of the
$m$-phase solutions. The parameters $\bm{\theta}_{0}$
represent the ``initial phase shifts'', which can take arbitrary
values on the family of the $m$-phase solutions.

 Let us say also here that the function $f ({\bf x})$ represents
a quasiperiodic function on $\mathbb{R}^{d}$ with the wave numbers
$({\bf k}_{1}, \dots, {\bf k}_{d})$ if it comes from a smooth
function $f (\bm{\theta})$ on the torus $\mathbb{T}^{m}$:
$$f ({\bf k}_{1} x^{1} + \dots + {\bf k}_{d} x^{d} +
\bm{\theta}_{0})  \,\,\,\,\, \rightarrow \,\,\,\,\,
f (x^{1}, \dots, x^{d}) $$
under the corresponding mapping
$\mathbb{R}^{d} \rightarrow \mathbb{T}^{m}$.

 Let us call a smooth family of $m$-phase solutions of (\ref{InSyst})
any family (\ref{mphasesol}) with a smooth dependence of the functions
$\bm{\Phi} (\bm{\theta}, {\bf U})$ on some finite number of parameters
${\bf U} = (U^{1}, \dots, U^{N})$.

 Here, however, we will need to generalize the definition of $m$-phase
solutions of system (\ref{InSyst}) to include the presence of the
pseudo-phases in the consideration. Let us say that the method of
pseudo-phases was introduced by Whitham in \cite{whith3} in connection
with the Lagrangian structure of the Whitham system for the
Korteweg - de Vries (KdV) equation. In this paper the appearance of
pseudo-phases will be connected with the geometrical or physical
meaning of the field variables $(\varphi^{1}, \dots, \varphi^{n})$.
Namely, it appears quite often that a part of the variables
$(\varphi^{1}, \dots, \varphi^{n})$ represents in fact some geometrical
(or physical) ``phase'' variables growing linearly with the spatial
or time variables. Thus, we have to separate the variables
$(\varphi^{1}, \dots, \varphi^{n})$ into two parts
\begin{equation}
\label{TwoVarType}
\left( \varphi^{1}, \dots, \varphi^{n} \right) \,\,\, = \,\,\,
\left( \rho^{1}, \dots, \rho^{n_{1}}, \, \phi^{1}, \dots, \phi^{n_{2}}
\right) \,\,\, , \,\,\,\,\,\,\,\, (n_{1} + n_{2} = n) \,\,\, ,
\end{equation}
representing the ``density-type'' and the ``phase-type''
variables respectively. Now, we will put slightly different conditions
for the variables $(\rho^{1}, \dots, \rho^{n_{1}})$ and
$(\phi^{1}, \dots, \phi^{n_{2}})$ in the  definition of the $m$-phase
solutions of (\ref{InSyst}) putting
\begin{equation}
\label{RhoRepr}
\rho^{i} ({\bf x}, t) \,\, = \,\, R^{i} \left(
{\bf k}_{1}({\bf U})\, x^{1} \, + \, \dots \, + \,
{\bf k}_{d}({\bf U})\, x^{d}
\, + \, \bm{\omega}({\bf U})\, t \, + \, \bm{\theta}_{0}, \,
{\bf U} \right) \,\,\,\,\, , \,\,\,\,\,\,\,\,\,\, i = 1, \dots, n_{1}
\end{equation}
\begin{multline}
\label{PhiRepr}
\phi^{j} ({\bf x}, t) \,\, = \,\, \Psi^{j} \left(
{\bf k}_{1}({\bf U})\, x^{1} \, + \, \dots \, + \,
{\bf k}_{d}({\bf U})\, x^{d}
\, + \, \bm{\omega}({\bf U})\, t \, + \, \bm{\theta}_{0}, \,
{\bf U} \right) \,\, + \\
+ \,\, p^{j}_{1} ({\bf U})\, x^{1} \, + \, \dots \, + \,
p^{j}_{d} ({\bf U})\, x^{d} \, + \,
\Omega^{j} ({\bf U})\, t \,\, + \,\, \tau^{j}_{0}
\,\,\,\,\, , \,\,\,\,\,\,\,\,\,\, j = 1, \dots, n_{2}
\end{multline}
with some $2\pi$-periodic in each $\theta^{\alpha}$ functions
${\bf R} (\bm{\theta}, {\bf U}) $,
$\, \bm{\Psi} (\bm{\theta}, {\bf U})$.

 It is natural to put also the normalization
\begin{equation}
\label{PsijNormCond}
\int_{0}^{2\pi}\!\!\!\dots\int_{0}^{2\pi}
\Psi^{j} \left( \bm{\theta} + \bm{\theta}_{0}, \, {\bf U} \right)
\, {d^{m} \theta \over (2\pi)^{m}} \,\,\, \equiv \,\,\, 0
\,\,\,\,\, , \,\,\,\,\,\,\,\,\,\, j = 1, \dots, n_{2}
\end{equation}

 According to the meaning of the variables
$\phi^{j} ({\bf x}, t)$, only their spatial or time derivatives
have in fact the physical sense, so, the right-hand part of system
(\ref{InSyst}) in the variables $(\bm{\rho}, \bm{\phi})$ should
contain only the spatial derivatives of $\phi^{j} ({\bf x}, t)$.
We can rewrite then the initial system (\ref{InSyst}) in the
variables $(\bm{\rho}, \bm{\phi})$  in the form
\begin{equation}
\label{rhophisystem}
\begin{array}{c}
\rho^{i}_{t} \,\, = \,\, A^{i} \left( \bm{\rho}, \bm{\rho}_{\bf x},
\bm{\phi}_{\bf x}, \bm{\rho}_{\bf xx}, \bm{\phi}_{\bf xx}, \dots
\right) \,\,\,\,\, , \,\,\,\,\,\,\,\,\,\, i = 1, \dots, n_{1} \\  \\
\phi^{j}_{t} \,\, = \,\, B^{j} \left( \bm{\rho}, \bm{\rho}_{\bf x},
\bm{\phi}_{\bf x}, \bm{\rho}_{\bf xx}, \bm{\phi}_{\bf xx}, \dots
\right) \,\,\,\,\, , \,\,\,\,\,\,\,\,\,\, j = 1, \dots, n_{2}
\end{array}
\end{equation}

 The functions $R^{i} (\bm{\theta}, {\bf U})$ and
$\Psi^{j} (\bm{\theta}, {\bf U})$ are then defined by the system
\begin{equation}
\label{RPsimphase}
\begin{array}{c}
\omega^{\alpha} R^{i}_{\theta^{\alpha}} \, - \,
A^{i} \left( {\bf R}, \,
k_{1}^{\beta_{1}} {\bf R}_{\theta^{\beta_{1}}},
\dots, k_{d}^{\beta_{d}} {\bf R}_{\theta^{\beta_{d}}}, \,
k_{1}^{\gamma_{1}} \bm{\Psi}_{\theta^{\gamma_{1}}} + {\bf p}_{1},
\dots
\right) \, = \, 0    \\   \\
\Omega^{j}  +  \omega^{\alpha} \Psi^{j}_{\theta^{\alpha}} \, - \,
B^{j} \left( {\bf R}, \,
k_{1}^{\beta_{1}} {\bf R}_{\theta^{\beta_{1}}},
\dots, k_{d}^{\beta_{d}} {\bf R}_{\theta^{\beta_{d}}}, \,
k_{1}^{\gamma_{1}} \bm{\Psi}_{\theta^{\gamma_{1}}} + {\bf p}_{1},
\dots  \right) \, = \, 0
\end{array}
\end{equation}
(summation over repeated indexes) with normalization conditions
(\ref{PsijNormCond}).

 In this paper we will need in fact to put more special requirements
to the definition of the pseudo-phases in the general Whitham scheme.
In particular, we will assume in this paper that the values
$({\bf k}_{1}, \dots, {\bf k}_{d}, \, \bm{\omega},
{\bf p}_{1}, \dots, {\bf p}_{d}, \, \bm{\Omega})$ represent
independent parameters on the family $\Lambda$ of $m$-phase
solutions of (\ref{rhophisystem}) such that the number of the
parameters ${\bf U}$ on $\Lambda$ is not less than \linebreak
$m(1+d) + n_{2}(1+d)$. Thus, we will assume here that the family
$\Lambda$ has $N = m(1+d) + n_{2}(1+d) + s$, $(s \geq 0)$
parameters except the initial phase shifts,
which can be chosen in the form
$$(U^{1}, \dots, U^{N}) \,\, = \,\,
({\bf k}_{1}, \dots, {\bf k}_{d}, \, \bm{\omega},
{\bf p}_{1}, \dots, {\bf p}_{d}, \, \bm{\Omega},
n^{1}, \dots, n^{s})$$
where $(n^{1}, \dots, n^{s})$ are some
additional parameters in the set $(U^{1}, \dots, U^{N})$ (if any).
In general, the parameters $(U^{1}, \dots, U^{N})$ can be chosen
in different ways, we just assume that they do not change under
the initial phase shifts on $\Lambda$. The parameters
$\tau^{j}_{0}$, $j = 1, \dots, n_{2}$, as well as
$\theta_{0}^{\alpha}$, $\alpha = 1, \dots, m$ represent the
initial phase shifts on the family $\Lambda$.

 Another important requirement on the pseudo-phases in our scheme
will be considered in the next chapter and is connected with
the Hamiltonian structure of system (\ref{InSyst}).

 As it is well known, in the Whitham approach
(\cite{whith1,whith2,whith3}) the parameters $(U^{1}, \dots, U^{N})$
become ``slow'' functions of coordinates and time. More precisely,
we have to make the coordinate change
$x^{q} \, \rightarrow \, X^{q} = \epsilon x^{q}$,
$t \, \rightarrow \, T = \epsilon t$, $\epsilon \rightarrow 0$ and
introduce the slow functions $S^{\alpha} ({\bf X}, T)$,
$\alpha = 1, \dots, m$, $\, \Sigma^{j} ({\bf X}, T)$,
$j = 1, \dots, n_{2}$. We try to construct then the asymptotic
solutions of the system
\begin{equation}
\label{epsrhophisystem}
\begin{array}{c}
\epsilon \, \rho^{i}_{T} \,\, = \,\,
A^{i} \left( \bm{\rho}, \, \epsilon \bm{\rho}_{\bf X}, \,
\epsilon \bm{\phi}_{\bf X}, \, \epsilon^{2} \bm{\rho}_{\bf XX}, \,
\epsilon^{2} \bm{\phi}_{\bf XX}, \, \dots
\right) \,\,\,\,\, , \,\,\,\,\,\,\,\,\,\, i = 1, \dots, n_{1} \\  \\
\epsilon \, \phi^{j}_{T} \,\, = \,\, B^{j} \left( \bm{\rho}, \,
\epsilon \bm{\rho}_{\bf X}, \, \epsilon \bm{\phi}_{\bf X}, \,
\epsilon^{2} \bm{\rho}_{\bf XX}, \,
\epsilon^{2} \bm{\phi}_{\bf XX}, \, \dots
\right) \,\,\,\,\, , \,\,\,\,\,\,\,\,\,\, j = 1, \dots, n_{2}
\end{array}
\end{equation}
with the main term having the form
\begin{multline}
\label{rho0phi0}
\rho_{(0)}^{i} \,\,\, = \,\,\, R^{i} \left(
{{\bf S} ({\bf X}, T) \over \epsilon} \, + \,
\bm{\theta}_{0}({\bf X}, T)
\, + \, \bm{\theta}, \,\, {\bf U} ({\bf X}, T) \! \right)
\,\,\, , \,\,\,\,\,\,\,\, i = 1, \dots, n_{1}  \, , \\
\phi_{(0)}^{j} \, = \,\, \Psi^{j} \left(
{{\bf S} ({\bf X}, T) \over \epsilon} \, + \,
\bm{\theta}_{0}({\bf X}, T)
\, + \, \bm{\theta}, \,\, {\bf U} ({\bf X}, T) \! \right)
\,\, + \,\, {1 \over \epsilon} \,
\Sigma^{j} ({\bf X}, T) \,\, + \,\,
\tau_{0}^{j} ({\bf X}, T) \,\,\,\,\, , \,\,\,\,\,\,\,\,  
j = 1, \dots, n_{2} \, .
\end{multline}

 Substituting the functions from $\Lambda$ it is easy to get the
relations
$$S^{\alpha}_{T} \,\, = \,\, \omega^{\alpha} ({\bf U})
\,\,\, , \,\,\,\,\,
S^{\alpha}_{X^{q}} \,\, = \,\, k^{\alpha}_{q} ({\bf U})
\,\,\, , \,\,\,\,\,
\Sigma^{j}_{T} \,\, = \,\, \Omega^{j} ({\bf U})
\,\,\, , \,\,\,\,\,
\Sigma^{j}_{X^{q}} \,\, = \,\, p^{j}_{q} ({\bf U}) $$
in the zero approximation, which gives the compatibility
conditions
\begin{equation}
\label{CompRel}
k^{\alpha}_{q \, T} \,\, = \,\, \omega^{\alpha}_{X^{q}}
\,\,\, , \,\,\,\,\,
p^{j}_{q \, T} \,\, = \,\, \Omega^{j}_{X^{q}}
\,\,\, , \,\,\,\,\,
k^{\alpha}_{q \, X^{p}} \,\, = \,\, k^{\alpha}_{p \, X^{q}}
\,\,\, , \,\,\,\,\,
p^{j}_{q \, X^{k}} \,\, = \,\, p^{j}_{k \, X^{q}}
\end{equation}
for the parameters
$({\bf k}_{1}, \dots, {\bf k}_{d}, \, \bm{\omega},
{\bf p}_{1}, \dots, {\bf p}_{d}, \, \bm{\Omega})$
on the family $\Lambda$.

 The second part of restrictions on the parameters
$(U^{1}, \dots, U^{N})$ in the Whitham method is given by the   
requirement of the existence of the first correction
$(\bm{\rho}_{(1)}, \, \bm{\phi}_{(1)})$ to solution (\ref{rho0phi0})
$$\rho^{i} \,\, \simeq \,\, \rho^{i}_{(0)} \,\, + \,\,
\epsilon \, \rho^{i}_{(1)} \left(
{{\bf S} ({\bf X}, T) \over \epsilon}
\, + \, \bm{\theta}, \,\, {\bf X}, \, T \right) $$
$$\phi^{i} \,\, \simeq \,\, \phi^{i}_{(0)} \,\, + \,\,
\epsilon \, \phi^{i}_{(1)} \left(
{{\bf S} ({\bf X}, T) \over \epsilon}
\, + \, \bm{\theta}, \,\, {\bf X}, \, T \right) $$
on the space of $2\pi$-periodic in $\bm{\theta}$ functions
(see \cite{luke}).

 The functions
$(\bm{\rho}_{(1)} (\bm{\theta}, {\bf X}, T), \,
\bm{\phi}_{(1)} (\bm{\theta}, {\bf X}, T))$ are defined by the
linear system
$${\hat L}_{[{\bf U}({\bf X}, T), \, \bm{\theta}_{0}({\bf X}, T)]}
\, \left(
\begin{array}{c}
\bm{\rho}_{(1)} (\bm{\theta}, {\bf X}, T) \cr
\bm{\phi}_{(1)} (\bm{\theta}, {\bf X}, T)
\end{array}
\right)
\,\,\, = \,\,\, {\bf f}_{1} (\bm{\theta}, {\bf X}, T) $$
where ${\hat L}_{[{\bf U}({\bf X}, T), \, \bm{\theta}_{0}({\bf X}, T)]}$
is the linear operator given by the linearization of the left-hand
part of system (\ref{RPsimphase}) on the corresponding functions from
$\Lambda$ and ${\bf f}_{1} (\bm{\theta}, {\bf X}, T)$ is the first
$\epsilon$-discrepancy defined after the substitution of
(\ref{rho0phi0}) in (\ref{epsrhophisystem}).

 The operator
${\hat L}_{[{\bf U}({\bf X}, T), \, \bm{\theta}_{0}({\bf X}, T)]}$
represents a differential in $\bm{\theta}$ operator with periodic
coefficients at every fixed ${\bf X}$ and $T$. We get then that the
second part of the Whitham system should be given by the orthogonality
of the function ${\bf f}_{1} (\bm{\theta}, {\bf X}, T)$ to all the
left eigen-vectors of ${\hat L}$ (the eigen-vectors of the adjoint
operator) corresponding to the zero eigen-values at every fixed
$({\bf X}, T)$.

 We should say, however, that the orthogonality of
${\bf f}_{1} (\bm{\theta}, {\bf X}, T)$ to all the left
eigen-vectors of ${\hat L}$ with zero eigen-values is imposed usually
just in the one-phase situation. In this case we have usually just
a finite number of such eigen-vectors depending regularly on the
parameters $(U^{1}, \dots, U^{N})$. The corresponding orthogonality
conditions together with conditions (\ref{CompRel})
give then a regular system of hydrodynamic type which represents the
Whitham system in the one-phase situation. Another important thing
taking place in the one-phase situation is the possibility of
constructing of all the corrections $\bm{\varphi}_{(n)}$ in all
orders of $\epsilon$ and representing the asymptotic solution as a
regular series in integer powers of $\epsilon$.

 This situation, however, does not usually take place in the
multi-phase case where the behavior of the eigen-vectors of ${\hat L}$
is usually much more complicated. Thus, the kernels of the operators
${\hat L}$ and ${\hat L}^{\dagger}$ depend usually in highly nontrivial
way on the parameters ${\bf U}$, being finite- or infinite-dimensional
for different values of $(U^{1}, \dots, U^{N})$. In this situation it
is natural to define the ``regular'' orthogonality conditions just by
the requirement of orthogonality of ${\bf f}_{1}$ to the ``regular''
set of the kernel vectors of ${\hat L}^{\dagger}$ which is usually
finite also in the multi-phase case. Thus, we assume here that the
kernels of the operators ${\hat L}$ and ${\hat L}^{\dagger}$
contain just a finite number of linearly independent ``regular''
eigen-vectors, i.e. the eigen-vectors smoothly depending on the
parameters ${\bf U}$. The ``regular'' Whitham system is defined in
this situation by conditions (\ref{CompRel})
and the orthogonality of the discrepancy
${\bf f}_{1} (\bm{\theta}, {\bf X}, T)$ to all the regular left
eigen-vectors of ${\hat L}$ corresponding to the zero eigen-value.

 Let us say that the first correction $\bm{\varphi}_{(1)}$ to the
asymptotic solution (\ref{rho0phi0}) can not be found here in such
a simple form as in the one-phase situation. However, as the
investigations of this situation show, the corrections to the main
approximation $\bm{\varphi}_{(0)}$ still vanish as
$\epsilon \rightarrow 0$ even in the multi-phase case
(see \cite{dobr1,dobr2,DobrKrichever}). So, despite the high
non-triviality of the next approximation in this case
(\cite{dobr1,dobr2,DobrKrichever}), the regular Whitham system
still plays very important role in consideration of slow-modulated
$m$-phase solutions.

 It is not difficult to see that the Whitham system imposes
restrictions just on the functions ${\bf U} ({\bf X}, T)$ and does
not contain the parameters $\bm{\theta}_{0} ({\bf X}, T)$ and
$\bm{\tau}_{0} ({\bf X}, T)$. Indeed, the functions
$\bm{\theta}_{0} ({\bf X}, T)$ and $\bm{\tau}_{0} ({\bf X}, T)$
can be considered just as $\epsilon$-corrections to the functions
${\bf S} ({\bf X}, T)$ and $\bm{\Sigma} ({\bf X}, T)$, so the
constraints arising on the first step include just the main terms
${\bf S} ({\bf X}, T)$ and $\bm{\Sigma} ({\bf X}, T)$, while the
restrictions on $\bm{\theta}_{0} ({\bf X}, T)$ and
$\bm{\tau}_{0} ({\bf X}, T)$ arise in the higher approximations
(if they exist) (see \cite{luke}).\footnote{A more detailed
discussion of the phase shift $\bm{\theta}_{0} ({\bf X}, T)$
can be found for example in
\cite{Haberman1, Haberman2,MaltsevJMP, DobrMinenkov}. We should
note also that the phase shift can play rather important role
in the weakly nonlinear case, leading to nontrivial corrections
to the Whitham system
(\cite{Newell}, see also \cite{MaltsevAmerMath, DobrMinenkov}).}

 For the correct construction of the modulated solutions and a
good definition of the Whitham system we have to require in fact
one more thing from the family $\Lambda$. Namely, the correct
procedure of constructing of modulated solutions can be
implemented on the ``complete regular families'' $\Lambda$ of
$m$-phase solutions of (\ref{rhophisystem}). Let us give here
the corresponding definition.  Let us consider the set of
parameters ${\bf U}$ in the form
$${\bf U} = ({\bf k}_{1}, \dots, {\bf k}_{d}, \, \bm{\omega},
{\bf p}_{1}, \dots, {\bf p}_{d}, \, \bm{\Omega},
n^{1}, \dots, n^{s})$$

 It is easy to see then that the vectors
$$\bm{\xi}_{(\alpha)[{\bf U}, \, \bm{\theta}_{0}]} \,\, = \,\,
\left( {\bf R}_{\theta^{\alpha}} (\bm{\theta} + \bm{\theta}_{0},
\, {\bf U}), \,\, \bm{\Psi}_{\theta^{\alpha}}
(\bm{\theta} + \bm{\theta}_{0}, \, {\bf U}) \right)^{t} \,\,\, ,
\,\,\,\,\,\,\,\, \alpha = 1, \dots, m \, , $$
$$\bm{\eta}_{(l)[{\bf U}, \, \bm{\theta}_{0}]} \,\, = \,\,
\left( {\bf R}_{n^{l}} (\bm{\theta} + \bm{\theta}_{0},
\, {\bf U}), \,\, \bm{\Psi}_{n^{l}}
(\bm{\theta} + \bm{\theta}_{0}, \, {\bf U}) \right)^{t} \,\,\, ,
\,\,\,\,\,\,\,\, l = 1, \dots, s \, , $$
$${\rm and} \,\,\,\,\,
\bm{\zeta}_{(j)[{\bf U}, \, \bm{\theta}_{0}]} \,\, = \,\,
\left( 0, \dots, 1, \dots, 0 \right)^{t} \,\,\,\,\,\,\,\,
((n_{1} + j){\rm th \,\,\, position}) \,\,\, ,
\,\,\,\,\,\,\,\, j = 1, \dots, n_{2} $$
represent regular (right) eigen-vectors of the operators
${\hat L}_{[{\bf U}, \, \bm{\theta}_{0}]}$ corresponding to the
zero eigen-value.

\vspace{0.2cm}

{\bf Definition 1.1.}

{\it We call family $\Lambda$ a complete regular family of
$m$-phase solutions of (\ref{rhophisystem}) with $n_{2}$
pseudo-phases if:

1) The values ${\bf k}_{p} = (k^{1}_{p}, \dots, k^{m}_{p})$,
$\bm{\omega} = (\omega^{1}, \dots, \omega^{m})$,
${\bf p}_{q} = (p_{q}^{1}, \dots, p_{q}^{n_{2}})$, and
$\bm{\Omega} = (\Omega^{1}, \dots, \Omega^{n_{2}})$
are all independent, such that the total set of independent
parameters on $\Lambda$ can be represented in the form
$$\left( {\bf U}, \, \bm{\theta}_{0}, \,
\bm{\tau}_{0} \right) \, = \, \left(
{\bf k}_{1}, \dots, {\bf k}_{d}, \, \bm{\omega}, \,
{\bf p}_{1}, \dots, {\bf p}_{d}, \, \bm{\Omega},
n^{1}, \dots, n^{s}, \, \bm{\theta}_{0}, \, \bm{\tau}_{0} \right)$$

2) The vectors $\bm{\xi}_{(\alpha)[{\bf U}, \, \bm{\theta}_{0}]}$,
$\bm{\eta}_{(l)[{\bf U}, \, \bm{\theta}_{0}]}$, and
$\bm{\zeta}_{(j)[{\bf U}, \, \bm{\theta}_{0}]}$ are linearly
independent and represent the maximal linearly independent set of
the kernel vectors of ${\hat L}_{[{\bf U}, \, \bm{\theta}_{0}]}$
smoothly depending on the parameters ${\bf U}$;

3) The operator ${\hat L}_{[{\bf U}, \, \bm{\theta}_{0}]}$ also has
exactly $m + s + n_{2}$ linearly independent left eigen-vectors
corresponding to the zero eigen-value
$$\bm{\kappa}^{(q)}_{[{\bf U}]} (\bm{\theta} + \bm{\theta}_{0})
\,\,\, = \,\,\,
\bm{\kappa}^{(q)}_{[{\bf k}_{1}, \dots, {\bf k}_{d}, \, \bm{\omega}, 
\, {\bf p}_{1}, \dots, {\bf p}_{d}, \, \bm{\Omega}, \, {\bf n}]}
(\bm{\theta} + \bm{\theta}_{0}) \,\,\, , \,\,\,\,\,
q = 1, \dots, m + s + n_{2} \,\,\, , $$
defined for all values of ${\bf U}$ and depending smoothly on 
the parameters ${\bf U}$.

}

\vspace{0.2cm}

 Let us call the regular Whitham system for a complete
regular family of $m$-phase solutions of system (\ref{rhophisystem}) with
$n_{2}$ pseudo-phases the conditions of orthogonality of the
discrepancy ${\bf f}_{(1)}(\bm{\theta}, {\bf X}, T)$ to the vectors
$\bm{\kappa}^{(q)}_{[{\bf U}({\bf X},T)]}
(\bm{\theta}\, + \, \bm{\theta}_{0}({\bf X},T))$
\begin{equation}
\label{ortcond}
\int_{0}^{2\pi}\!\!\!\!\!\dots\int_{0}^{2\pi}
\kappa^{(q)}_{[{\bf U}({\bf X},T)]\, i}
(\bm{\theta}\, + \, \bm{\theta}_{0}({\bf X},T)) \,\,
f^{i}_{(1)} (\bm{\theta},{\bf X},T) \,\,
{d^{m} \theta \over (2\pi)^{m}} \,\,\, = \,\,\, 0
\end{equation}
($q = 1, \, \dots , \, m + s + n_{2}$)
and the compatibility conditions
\begin{equation}
\label{TimeConstraints}
k^{\alpha}_{p \, T} \,\, = \,\, \omega^{\alpha}_{X^{p}}
\,\,\,\,\,\,\,\, , \,\,\,\,\,\,\,\,
p^{j}_{l \, T} \,\, = \,\, \Omega^{j}_{X^{l}}
\end{equation}
\begin{equation}
\label{SpatialConstraints}
k^{\alpha}_{p \, X^{l}} \,\, = \,\, k^{\alpha}_{l \, X^{p}}
\,\,\,\,\,\,\,\, , \,\,\,\,\,\,\,\,
p^{j}_{l \, X^{k}} \,\, = \,\, p^{j}_{k \, X^{l}}
\end{equation}
$\alpha = 1, \dots, m$, $\,\, p, l, k = 1, \dots, d$,
$\,\, j = 1, \dots, n_{2}$.

 For our further purposes it will be convenient to separate
the evolutionary part of the Whitham system and purely spatial
constraints. So, let us call here relations
(\ref{ortcond}) - (\ref{TimeConstraints}) the evolutionary part of
a regular Whitham system. The relations (\ref{SpatialConstraints}) 
will be considered then as additional constraints for the evolutionary 
system (\ref{ortcond}) - (\ref{TimeConstraints}).

 The evolutionary part of a regular Whitham system provides
exactly \linebreak
$m (d + 1) + n_{2} (d + 1) + s$ independent relations
for $N = m (d + 1) + n_{2} (d + 1) + s$ 
parameters
${\bf U} = ({\bf k}_{1}, \dots, {\bf k}_{d}, \, \bm{\omega},
{\bf p}_{1}, \dots, {\bf p}_{d}, \, \bm{\Omega}, \, {\bf n})$
at every ${\bf X}$ and $T$. We can assume also, that in generic case 
the derivatives ${\bf U}_{T}$ can be expressed in terms of the 
spatial derivatives ${\bf U}_{X^{l}}$, such that we can write the 
evolutionary part of a regular Whitham system in the form
\begin{equation}
\label{HTSystem}
U^{\nu}_{T} \,\,\, = \,\,\, V^{\nu l}_{\mu} \left( {\bf U} \right)
\,\, U^{\mu}_{X^{l}}
\end{equation}
for general set of parameters ${\bf U}$.

 Following B.A. Dubrovin and S.P. Novikov we will call systems
having the form (\ref{HTSystem}) the systems of Hydrodynamic Type
in $d$ spatial dimensions.

 The Hamiltonian theory of systems (\ref{HTSystem}) was started
by B.A. Dubrovin and S.P. Novikov who introduced the concept of
the Poisson bracket of Hydrodynamic Type
(\cite{dn1,DubrNovDAN84,dn2,dn3}). The local Poisson brackets of
Hydrodynamic Type (Dubrovin - Novikov brackets) can be represented
by the following general form
\begin{equation}
\label{DNbracket}
\left\{ U^{\nu} ({\bf X}), U^{\mu} ({\bf Y}) \right\} \,\, = \,\,
g^{\nu\mu \, l} \left( {\bf U}({\bf X}) \right) \,
\delta_{X^{l}} ( {\bf X} - {\bf Y} ) \,\, + \,\,
b^{\nu\mu \, l}_{\lambda} \left( {\bf U}({\bf X}) \right) \,
U^{\lambda}_{X^{l}} \, \delta ( {\bf X} - {\bf Y} )
\end{equation}
(summation over repeated indexes).

 The theory of brackets (\ref{DNbracket}) is
best developed in the case of one spatial ($d = 1$) dimension.
Thus, expression (\ref{DNbracket}) with non-degenerate tensor
$g^{\nu\mu}$ defines a Poisson bracket for $d = 1$ if and only if
the tensor $g^{\nu\mu} ({\bf U})$ represents a flat pseudo-Riemannian
(contravariant) metric on the space of parameters ${\bf U}$,
while the functions
$\Gamma^{\nu}_{\mu\gamma} ({\bf U}) = - \, g_{\mu\lambda} ({\bf U})
\, b^{\lambda\nu}_{\gamma} ({\bf U}) \,\,\,$
($g^{\nu\lambda} ({\bf U}) \, g_{\lambda\mu} ({\bf U}) \, \equiv \,
\delta^{\nu}_{\mu}$) represent the corresponding Christoffel symbols.
As a corollary, every Dubrovin - Novikov bracket in one-dimensional
case can be written in the canonical (constant) form
$$\left\{ c^{\nu} (X) \, , \, c^{\mu} (Y) \right\} \,\, = \,\,
e^{\nu} \, \delta^{\nu\mu} \, \delta^{\prime} ( X -  Y )
\,\,\, , \,\,\,\,\, e^{\nu} = \pm 1 $$
using the flat coordinates $c^{\nu} = c^{\nu} ({\bf U})$ 
of the metric $g_{\nu\mu} ({\bf U})$.

 It's not difficult to see also that the functionals
$$C^{\nu} \,\, = \,\, \int_{-\infty}^{+\infty}
c^{\nu}(X) \,\, dX \,\,\,\,\, , \,\,\,\,\,\,\,\,
P \,\, = \,\, \int_{-\infty}^{+\infty} {1 \over 2}
\sum_{\nu=1}^{N} e^{\nu}  \, (c^{\nu})^{2} (X) \,\, dX$$
represent annihilators and the momentum functional of
bracket (\ref{DNbracket}) for the case $d =1$.
The systems of Hydrodynamic Type are generated by the functionals
of Hydrodynamic Type
$$H  \,\,\, = \,\,\,  \int_{-\infty}^{+\infty}
h({\bf U}) \,\, dX $$
according to the Dubrovin - Novikov bracket.

 The Hamiltonian approach plays very important role in the
theory of integrability of the Hydrodynamic Type systems in the
case of one spatial dimension. Thus, according to the conjecture 
of S.P. Novikov, any system of Hydrodynamic Type which can be
written in the diagonal form
$$U^{\nu}_{T} \,\,\, = \,\,\, V^{\nu} \left( {\bf U} \right)
\, U^{\nu}_{X}$$
and is Hamiltonian with respect to some local bracket of
Hydrodynamic Type is integrable. The Novikov conjecture was
proved by S.P. Tsarev (\cite{tsarev,tsarev2}), who also suggested a
method of integration of systems of this kind. The method suggested 
by Tsarev (the generalized hodograph method) can be applied in fact
to a wider class of ``semi-Hamiltonian'' systems, which contains all 
the  diagonalizable Hamiltonian systems as a subclass. As was shown 
later, the class of ``semi-Hamiltonian systems'' contains also the
systems, Hamiltonian with respect to the Mokhov - Ferapontov bracket 
(\cite{mohfer1}) or the Ferapontov brackets (\cite{fer1, fer2}), 
which can be considered as the weakly nonlocal generalizations 
of the Dubrovin - Novikov bracket. Let us give here the references on 
papers \cite{mohfer1, fer1, fer2, fer3, fer4, Pavlov1, PhysD}
where the detailed discussion of the weakly nonlocal Poisson
structures can be found.

 Let us say, that the theory of the Dubrovin - Novikov
brackets in the multi-dimensional case is more complicated
than in the case $d = 1$.
The most general properties of the multi-dimensional
brackets (\ref{DNbracket}) were investigated in
\cite{DubrNovDAN84,MokhovMultDimBr1,MokhovMultDimBr2}. However,
the investigation of the brackets (\ref{DNbracket}) in $d >1$
dimensions still represents one of the most interesting branch of
the theory of infinite-dimensional Poisson structures.

 The Hamiltonian formulation of the Whitham method was also
suggested by B.A. Dubrovin and S.P. Novikov who introduced the
procedure of ``averaging'' of Hamiltonian structures in the
theory of slow modulations (\cite{dn1,dn2,dn3}). This approach
is connected with the Whitham method for the evolutionary systems
$$\varphi^{i}_{t} \,\, = \,\, F^{i} (\bm{\varphi}, \bm{\varphi}_{x},
\dots ) $$
having a local field-theoretic Poisson structure
$$\{\varphi^{i}(x) , \varphi^{j}(y)\} \,\, = \,\, \sum_{k \geq 0}
\, B^{ij}_{(k)} (\bm{\varphi}, \bm{\varphi}_{x}, \dots)
\,\, \delta^{(k)}(x-y) $$
with the local Hamiltonian of the form
$$H \,\, = \,\, \int P_{H}
(\bm{\varphi}, \bm{\varphi}_{x}, \dots) \, dx $$

 The procedure of averaging of local field-theoretic Poisson
brackets was first developed in the case of one spatial dimension
and gives a local Poisson structure of Hydrodynamic Type for the
corresponding Whitham system. The method of B.A. Dubrovin and
S.P. Novikov is connected with the conservative form of the
Whitham system and is based on the existence of a set of 
commuting local integrals
$$I^{\nu} \,\, = \,\, \int
P^{\nu}(\bm{\varphi}, \bm{\varphi}_{x},\dots) \, dx $$
which number is equal to the number of parameters $U^{\nu}$ on the 
family $\Lambda$.

 The integrals $I^{\nu}$ should satisfy the relations
$$\{I^{\nu} , H\} = 0 \,\,\,\,\, , \,\,\,\,\,
\{I^{\nu} , I^{\mu}\} = 0 \,\,\, ,$$
such that we can write for the time evolution of the densities
$P^{\nu} (\bm{\varphi}, \bm{\varphi}_{x},\dots)$:
$$P^{\nu}_{t} (\bm{\varphi}, \bm{\varphi}_{x},\dots) \,\, \equiv \,\,
Q^{\nu}_{x} (\bm{\varphi}, \bm{\varphi}_{x},\dots) $$
with some functions $Q^{\nu} (\bm{\varphi}, \bm{\varphi}_{x},\dots)$.
In the same way, the calculation of the Poisson brackets
of the densities $P^{\nu}$ gives the relations
$$\{P^{\nu}(x) \, , \, P^{\mu}(y)\} \,\, = \,\, \sum_{k\geq 0}
A^{\nu\mu}_{k}(\bm{\varphi}, \bm{\varphi}_{x}, \dots) \,
\delta^{(k)}(x-y) $$
where
$$A^{\nu\mu}_{0}(\bm{\varphi}, \bm{\varphi}_{x}, \dots)
\,\, \equiv \,\,
\partial_{x} Q^{\nu\mu}(\bm{\varphi}, \bm{\varphi}_{x}, \dots) $$
with some local functions 
$Q^{\nu\mu}(\bm{\varphi}, \bm{\varphi}_{x}, \dots)$.

 It is natural to define the procedure
$\langle \dots \rangle$ of averaging of any expression
$f (\bm{\varphi}, \bm{\varphi}_{x}, \dots)$ over the phase
variables on $\Lambda$ putting
$$\langle f \rangle \,\, = \,\,
\int_{0}^{2\pi}\!\!\!\dots\int_{0}^{2\pi}
f \left( \bm{\Phi}, \, k^{\alpha} \bm{\Phi}_{\theta^{\alpha}},
\dots \right) \,\, {d^{m} \theta \over (2\pi)^{m}} $$

 The Dubrovin - Novikov bracket on the space of functions
${\bf U}({\bf X})$, where $U^{\nu} \equiv \langle P^{\nu} \rangle$,
is defined by the formula
\begin{equation}
\label{dubrnovb}
\{U^{\nu}(X) \, , \, U^{\mu}(Y)\} \,\, = \,\,
\langle A^{\nu\mu}_{1}\rangle
({\bf U}) \,\, \delta^{\prime}(X-Y) \,\, + \,\,
{\partial \langle Q^{\nu\mu} \rangle \over
\partial U^{\gamma}} \,\, U^{\gamma}_{X} \,\, \delta (X-Y)
\end{equation}

 The Whitham system can be written now in the form
$$\langle P^{\nu} \rangle_{T} \,\, = \,\,
\langle Q^{\nu} \rangle_{X} \,\,\,\,\, , \,\,\,\,\,\,\,\,
\nu = 1, \dots, N $$
and can be proved to be Hamiltonian with respect
to the bracket (\ref{dubrnovb}) with the
Hamiltonian of Hydrodynamic Type
$$H_{av} \,\, = \,\, \int_{-\infty}^{+\infty}
\langle P_{H} \rangle \left( {\bf U} (X) \right) \,\, d X $$

 The Jacobi identity for bracket (\ref{dubrnovb}) was first proved 
in \cite{izvestia} using some regularity assumptions about the 
family $\Lambda$. A more detailed consideration of the justification
of the Dubrovin - Novikov procedure in the single-phase and the
multi-phase situations was presented in \cite{DNMultDim}. In particular, 
it was first shown in \cite{DNMultDim} that the justification of the 
procedure can be done also in the presence of ``resonances'' which can 
arise in the multi-phase situation. Let us note, that in \cite{MalPav} 
it was shown also that the method of averaging of the Lagrangian
functional (\cite{whith3}) can be also considered in terms of the
Dubrovin - Novikov procedure for a wide class of local Lagrangian
systems. In \cite{malnloc2} the generalization of the
Dubrovin - Novikov procedure for the weakly nonlocal brackets
was also suggested.

 Let us say that the investigation of the Hamiltonian properties
of the Whitham systems was of a great interest since the pioneer
works of B.A. Dubrovin and S.P. Novikov. Besides that, the general
theory of the Dubrovin - Novikov brackets appeared to be extremely
important in many subjects. As the most striking example, we can
point out here the theory of the Frobenius manifolds built by B.A.
Dubrovin and based on the theory of compatible Dubrovin - Novikov
brackets (see e.g. \cite{Dubrovin1, DubrovinCommMathPhys, Dubrovin2}).

Among the papers devoted to the Hamiltonian structures of the
Whitham systems we would like to cite here also the papers
\cite{Pavlov2, Alekseev} where the local and the weakly
nonlocal Hamiltonian structures for the famous integrable    
hierarchies were considered.

 Unfortunately, we can not present here the complete list of papers
devoted to the Whitham approach. Let us just give here some incomplete 
list of classical papers where the fundamental aspects of the Whitham 
method were discussed \cite{AblBenny, dm, DobrMaslMFA, dobr1, dobr2, 
DobrKrichever, dn1, dn2, dn3, ffm, Hayes, krichev1, KricheverCPAM, luke, 
Newell, theorsol, Nov, whith1, whith2, whith3}. Let us say also, that
we will discuss here just the Hamiltonian properties of the 
multi-dimensional Whitham systems in the case of the presence of the 
pseudo-phases.

 In paper \cite{JMPMultDim} the procedure of averaging of
multi-dimensional local field-theoretic Poisson brackets was
suggested. The approach used in \cite{JMPMultDim} can be considered
as a generalization of the Dubrovin - Novikov procedure to the
multi-dimensional case. According to the approach of \cite{JMPMultDim}
we consider the regular Whitham system for a complete regular family
$\Lambda$ of $m$-phase solutions of system (\ref{InSyst}),
parametrized by the values
$({\bf k}_{1}, \dots, {\bf k}_{d}, \, \bm{\omega}, \, {\bf n}, \,
\bm{\theta}_{0})$. We assume now that system (\ref{InSyst}) is
Hamiltonian with respect to a local field-theoretic Poisson bracket
\begin{equation}
\label{MultDimPBr}
\{ \varphi^{i} ({\bf x}) \, , \, \varphi^{i} ({\bf y}) \} \,\, = \,\,
\sum_{l_{1},\dots,l_{d}} B^{ij}_{(l_{1},\dots,l_{d})}
(\bm{\varphi}, \bm{\varphi}_{\bf x}, \dots ) \,\,
\delta^{(l_{1})} (x^{1} - y^{1}) \, \dots \,
\delta^{(l_{d})} (x^{d} - y^{d})
\end{equation}
$(l_{1}, \dots, l_{d} \geq 0)$, with a local Hamiltonian of the form
\begin{equation}
\label{MultDimHamFunc}
H \,\, = \,\, \int P_{H} \left(\bm{\varphi}, \bm{\varphi}_{\bf x},
\bm{\varphi}_{\bf xx}, \dots \right) \,\, d^{d} x
\end{equation}

 Like in the Dubrovin - Novikov procedure we have to require here
the existence of $N$ (equal to the number of parameters
$({\bf k}_{1}, \dots, {\bf k}_{d}, \, \bm{\omega}, \, {\bf n})$)
first integrals
\begin{equation}
\label{Inuvarphi}
I^{\nu} \,\, = \,\, \int
P^{\nu} \left( \bm{\varphi}, \, \bm{\varphi}_{\bf x}, \,
\bm{\varphi}_{\bf xx}, \dots \right) \, d^{d} x
\end{equation}
such that their values can be chosen as the parameters
$(U^{1}, \dots, U^{N})$ on the family $\Lambda$. We assume also
that all the integrals $I^{\nu}$ commute with each
other and with the Hamiltonian $H$
\begin{equation}
\label{MultDimInvRel}
\{ I^{\nu} \, , \, I^{\mu} \} \, = \, 0
\,\,\,\,\, , \,\,\,\,\,
\{ I^{\nu} \, , \, H \} \, = \, 0
\end{equation}
according to bracket (\ref{MultDimPBr}). For the time evolution
of the densities $P^{\nu} ({\bf x})$ we can write
$$P^{\nu}_{t} \left( \bm{\varphi}, \bm{\varphi}_{\bf x},
\bm{\varphi}_{\bf xx}, \dots \right) \,\, = \,\,
Q^{\nu 1}_{x^{1}} \left( \bm{\varphi},
\bm{\varphi}_{\bf x}, \bm{\varphi}_{\bf xx}, \dots \right) \, + \,
\dots \, + \, Q^{\nu d}_{x^{d}} \left( \bm{\varphi},
\bm{\varphi}_{\bf x}, \bm{\varphi}_{\bf xx}, \dots \right) $$
with some functions $Q^{\nu l}$.

 In fact, we have to put also some additional requirements on the
family $\Lambda$ and the set of the integrals $I^{\nu}$. Namely,
we have to require that the family $\Lambda$ represents a regular
Hamiltonian family of $m$-phase solutions of system (\ref{InSyst})
and the set $(I^{1}, \dots, I^{N})$ represents a complete
Hamiltonian set of commuting integrals. So, we put in fact the 
following requirements:

\vspace{0.2cm}

1) The family $\Lambda$ represents a complete regular family of
$m$-phase solutions of system (\ref{InSyst}) according to 
Definition 1.1;

2) The bracket (\ref{MultDimPBr}) has the same number of 
annihilators $(N^{1}$, $\dots$, $N^{s})$ on the space of the
quasiperiodic functions with the wave numbers
$({\bf k}_{1}, \dots, {\bf k}_{d})$ for every fixed values of
$({\bf k}_{1}, \dots, {\bf k}_{d})$;

3) The values of the functionals $(I^{1}, \dots, I^{N})$
on the family $\Lambda$ represent the set of parameters 
$(U^{1}, \dots, U^{N})$ on this family;

4) The Hamiltonian flows, generated by the functionals
$(I^{1}, \dots, I^{N})$ according to bracket (\ref{MultDimPBr}),
generate on $\Lambda$ linear phase shifts of $\bm{\theta}_{0}$
with frequencies $\bm{\omega}^{\nu} ({\bf U})$, such that
$${\rm rk} \,\, || \omega^{\alpha \nu} ({\bf U}) || \,\, = \,\, m $$

5) At every ``point'' of the ``submanifold'' $\Lambda$, having
``coordinates'' 
$({\bf k}_{1}, \dots, {\bf k}_{d}, \bm{\omega}, {\bf n}, 
\bm{\theta}_{0})$, the linear space generated by the variation 
derivatives
$\delta I^{\nu} / \delta \varphi^{i} ({\bf x})$ contains
the variation derivatives of all the corresponding
annihilators $N^{q}$ of the bracket (\ref{MultDimPBr}), 
such that we can write
$$\left. {\delta N^{l} \over \delta \varphi^{i} ({\bf x})}
\right|_{\Lambda} \,\, = \,\,
\sum_{\nu=1}^{N} \gamma^{l}_{\nu} ({\bf U}) \,\, \left.
{\delta I^{\nu} \over \delta \varphi^{i} ({\bf x})} \right|_{\Lambda}$$
for some functions $\gamma^{l}_{\nu} ({\bf U})$ on $\Lambda$.

\vspace{0.2cm}

 Under the requirements formulated above the set
$(I^{1}, \dots, I^{N})$ can be used for construction of a local
field-theoretic Poisson bracket for the regular Whi\-tham system
on a regular Hamiltonian family $\Lambda$ of $m$-phase solutions
of system (\ref{InSyst}). The corresponding procedure in the
absence of the pseudo-phases can be formulated in the following
way:

 The pairwise Poisson brackets of the densities $P^{\nu} ({\bf x})$, 
$P^{\mu} ({\bf y})$ can be represented in the form
$$\{ P^{\nu} ({\bf x}) \, , \, P^{\mu} ({\bf y}) \} \,\, = \,\,
\sum_{l_{1},\dots,l_{d}} A^{\nu\mu}_{l_{1} \dots l_{d}}
(\bm{\varphi}, \bm{\varphi}_{\bf x}, \dots ) \,\,
\delta^{(l_{1})} (x^{1} - y^{1}) \, \dots \,
\delta^{(l_{d})} (x^{d} - y^{d}) $$
($l_{1}, \dots, l_{d} \geq 0$). In the same way as in the 
one-dimensional case, we can also write here the relations
$$A^{\nu\mu}_{0 \dots 0} (\bm{\varphi}, \bm{\varphi}_{\bf x}, \dots )
\,\,\, \equiv \,\,\, \partial_{x^{1}} \, Q^{\nu\mu 1}
(\bm{\varphi}, \bm{\varphi}_{\bf x}, \dots ) \, + \, \dots \, + \,
\partial_{x^{d}} \, Q^{\nu\mu d}
(\bm{\varphi}, \bm{\varphi}_{\bf x}, \dots ) $$
according to relations (\ref{MultDimInvRel}).
Let us say, however, that the averaged Poisson bracket does not have in
general the form (\ref{DNbracket}) for $d > 1$, which is connected with
the fact that the Hamiltonian structure should be defined now just
on the ``submanifold'' in the space of functions
${\bf U} ({\bf X})$, given by the constraints
$\, k^{\alpha}_{q \, X^{p}} \, = \, k^{\alpha}_{p \, X^{q}}$,
$\,\, \alpha = 1, \dots, m$, $\,\, q, p = 1, \dots, d$. To define
the corresponding Poisson bracket we have to introduce the
coordinates $S^{\alpha} ({\bf X})$ $\,\, (\alpha = 1, \dots, m)$
on this submanifold, defined by the relations
$S^{\alpha}_{X^{q}} = k^{\alpha}_{q} ({\bf X})$. It is easy to
see, that the spatial derivatives of the functions
$S^{\alpha} ({\bf X})$ provide just $m d$ coordinates on the
family $\Lambda$, connected with the wave numbers of the
solutions. For the remaining $m + s$ coordinates we can use just
arbitrary independent values $U^{\gamma}$,
$\,\, \gamma = 1, \dots, m + s$ from the full set
$U^{\nu} = \langle P^{\nu} \rangle$, $\,\, \nu = 1, \dots, N$
on $\Lambda$. The corresponding regular Whitham system on $\Lambda$
can then be written in the form:
\begin{equation}
\label{STOmegaRel}
\begin{array}{c}
S^{\alpha}_{T} \,\,\, = \,\,\, \omega^{\alpha}
\left({\bf S}_{\bf X}, \, U^{1}, \dots, U^{m+s} \right)
\,\,\,\,\, , \,\,\,\,\,\,\,\, \alpha = 1, \dots, m  \,\,\, ,
\\  \\
U^{\gamma}_{T} \,\,\, = \,\,\, \langle Q^{\gamma 1} \rangle_{X^{1}}
\,\, + \,\, \dots \,\, + \,\, \langle Q^{\gamma d} \rangle_{X^{d}}
\,\,\,\,\, , \,\,\,\,\,\,\,\, \gamma = 1, \dots, m + s \,\,\, ,
\end{array}
\end{equation}
where
$\langle Q^{\gamma p} \rangle \, = \, \langle Q^{\gamma p} \rangle
({\bf S}_{\bf X}, \, U^{1}, \dots, U^{m+s})$.

 It can be shown then that the Hamiltonian structure of system
(\ref{STOmegaRel}) is given by the Poisson bracket
\begin{equation}
\label{AveragedBracket}
\begin{array}{c}
\left\{ S^{\alpha} ({\bf X})  \, , \,
S^{\beta} ({\bf Y}) \right\} \,\, =  \,\, 0 \,\,\, ,   \\  \\
\left\{ S^{\alpha} ({\bf X})  \, , \,
U^{\gamma} ({\bf Y}) \right\} \,\, = \,\,
\omega^{\alpha\gamma}
\left({\bf S}_{\bf X},
U^{1}({\bf X}), \dots, U^{m+s}({\bf X}) \! \right)
\, \delta ({\bf X} - {\bf Y}) \,\,\, ,  \\  \\
\left\{ U^{\gamma} ({\bf X})\, , \, U^{\rho} ({\bf Y}) \right\}
\,\,\, =  \,\,\, \langle A^{\gamma\rho}_{10\dots0} \rangle
\left( {\bf S}_{\bf X}, \,
U^{1}({\bf X}), \dots, U^{m+s}({\bf X}) \right) \,\,
\delta_{X^{1}} ({\bf X} - {\bf Y}) \,\,\, +  \\   \\
+ \,\, \dots \,\, + \,\,\, \langle A^{\gamma\rho}_{0\dots01} \rangle
\left( {\bf S}_{\bf X}, \,
U^{1}({\bf X}), \dots, U^{m+s}({\bf X}) \right) \,\,\,
\delta_{X^{d}} ({\bf X} - {\bf Y}) \,\,\, +  \\   \\
+ \,\,\, \left[ \langle Q^{\gamma\rho \, p} \rangle
\left( {\bf S}_{\bf X}, \,
U^{1}({\bf X}), \dots, U^{m+s}({\bf X}) \right)
\right]_{X^{p}} \,\,\, \delta ({\bf X} - {\bf Y})
\,\,\,\,\, , \,\,\,\,\,\,\,\,
\gamma, \rho \, = \, 1, \dots , m + s \, ,
\end{array}
\end{equation}
with the Hamiltonian functional
$$H_{av} \,\,\, = \,\,\, \int \langle P_{H} \rangle \,
\left({\bf S}_{\bf X}, \, U^{1}({\bf X}), \dots, U^{m+s}({\bf X})
\right) \,\, d^{d} X $$

 Let us note here, that although just a part of the integrals
$I^{\nu}$ is formally used in the final construction of the
Hamiltonian structure, the presence of the complete Hamiltonian
set $(I^{1}, \dots, I^{N})$ plays an important role according to
the scheme of \cite{JMPMultDim}. The requirement of existence of
the complete set of local conservation laws is actually rather 
strong in the multi-dimensional ($d > 1$) situation. Thus, for
most of the integrable multi-dimensional systems the procedure,
formulated above, can not be used for general $m > 1$ since
only a finite set of local conservation laws is usually present
in this case. On the other hand, the procedure usually works well
in the single-phase $(m = 1$) case both in the integrable and
non-integrable situations.

 In this paper we are going to investigate the question if the
necessary number of the integrals $I^{\nu}$ can be reduced still
keeping the procedure of the bracket averaging well-justified.
As we will show, the number of the integrals $I^{\nu}$ can be
reduced in the case when a part of the phase variables
$\bm{\theta}_{0}$ can in fact be represented as the pseudo-phases.
As the analysis of different examples shows, this situation
actually takes place quite often. Moreover, in many cases the
existence of the multi-phase solutions for non-integrable
systems is caused in fact by the presence of the pseudo-phases, 
playing the role of additional phases of the solutions. Thus,
in many physical systems, the additional phases arise due to the
presence of some global additional symmetries, corresponding 
to the additional integrals of the system. The multi-phase
solutions can be considered in this case in fact as the periodic
waves with the parameters $(k_{1}, \dots, k_{d}, \omega)$ in
the nontrivial vacuum, while the additional parameters separate  
different vacua carrying permanent current. The simplest example
of such situation can be given just by the nonlinear Shr\"odinger 
equation with $d$ spatial dimensions, so we consider this example
at the end of the paper.

 In the next Chapter we will consider the procedure of the
bracket averaging in the presence of the pseudo-phases.

\section{The regularity conditions and the bracket averaging.}
\setcounter{equation}{0}

 As we said in the previous Chapter, we will consider here systems
(\ref{InSyst}) which can be represented in the Hamiltonian form
with some local field-theoretic Poisson bracket (\ref{MultDimPBr})
and the Hamiltonian functional (\ref{MultDimHamFunc}). Let us say,
that the space of fields $\bm{\varphi} ({\bf x})$ has a
pseudo-phase structure with $n_{2}$ pseudo-phases if we have a
(almost everywhere) free action of a $n_{2}$-dimensional Abelian
group $G^{n_{2}}$ on the target space
$(\varphi^{1}, \dots, \varphi^{n})$. We will say here, that the
pseudo-phase structure is compatible with the Poisson bracket
(\ref{MultDimPBr}) if the Poisson bracket is invariant under the
action of $G^{n_{2}}$.

 Easy to see that for the variables
$(\varphi^{1}, \dots, \varphi^{n})$, represented in the form
(\ref{TwoVarType}), the compatibility of the pseudo-phase structure
and the bracket (\ref{MultDimPBr}) means that the functions
$B^{ij}_{(l_{1}, \dots, l_{d})}$ depend just on the spatial
derivatives of the fields
$(\phi^{1}({\bf x}), \dots, \phi^{n_{2}}({\bf x}))$. We can see,
in particular, that the coefficients $B^{ij}_{(l_{1}, \dots, l_{d})}$
represent quasiperiodic functions on the family $\Lambda$
defined by formulas (\ref{RhoRepr}) - (\ref{PhiRepr}).

 We will say also here that the Hamiltonian system (\ref{InSyst})
is compatible with the pseudo-phase structure if both the bracket
(\ref{MultDimPBr}) and the Hamiltonian functional
(\ref{MultDimHamFunc}) are invariant under the action of the group
$G^{n_{2}}$.

 According to the scheme of the previous Chapter, we are going to
consider the Hamiltonian system (\ref{InSyst}), which is compatible
with the action of the Abelian group
\begin{equation}
\label{pseudophasegroup}
\begin{array}{c}
\left( \rho^{1}({\bf x}), \dots, \, \rho^{n_{1}}({\bf x}), \,\,
\phi^{1} ({\bf x}), \dots, \, \phi^{n_{2}}({\bf x}) \right)
\,\,\, \rightarrow   \\   \\
\rightarrow \,\,\,
\left( \rho^{1}({\bf x}), \dots, \, \rho^{n_{1}}({\bf x}), \,\,
\phi^{1} ({\bf x}) \, + \, \tau^{1}_{0}, \dots, \,
\phi^{n_{2}}({\bf x}) \, + \, \tau^{n_{2}}_{0} \right) \,\,\,\,\, ,
\end{array}
\end{equation}
and a complete regular family $\Lambda$ of $m$-phase solutions
of system (\ref{InSyst}) (or (\ref{rhophisystem})) with $n_{2}$
pseudo-phases, represented by relations
(\ref{RhoRepr}) - (\ref{PhiRepr}) and (\ref{PsijNormCond}).

 Let us define now a regular Hamiltonian family $\Lambda$
of $m$-phase solutions with $n_{2}$ pseudo-phases.

\vspace{0.2cm}

{\bf Definition 2.1.}

{\it We call family $\Lambda$ of $m$-phase solutions of system
(\ref{InSyst}) (or (\ref{rhophisystem})) with $n_{2}$ pseudo-phases
a regular Hamiltonian family if :

1) It represents a complete regular family of $m$-phase solutions
of system (\ref{InSyst}) (or (\ref{rhophisystem})) with $n_{2}$
pseudo-phases in the sense of Definition 1.1;

2) System (\ref{InSyst}) (or (\ref{rhophisystem})) represents
a Hamiltonian system compatible with the pseudo-phase structure
given by the action of the group (\ref{pseudophasegroup});

3) The Poisson bracket (\ref{MultDimPBr}) has on $\Lambda$
constant number of ``annihilators'' given by linearly
independent quasiperiodic solutions
$v^{(l)}_{i}({\bf x})$ of the system
$$\sum_{l_{1},\dots,l_{d}} \left.
B^{ij}_{(l_{1},\dots,l_{d})} (\bm{\varphi},
\bm{\varphi}_{\bf x}, \dots)
\right|_{\Lambda} v^{(l)}_{j, \, l_{1} x^{1} \dots \, l_{d} x^{d}}
({\bf x}) \,\,\, = \,\,\, 0 \,\,\,\,\, , $$
where $v^{(l)}_{i}({\bf x})$ have the same wave numbers
$({\bf k}_{1}, \dots, {\bf k}_{d})$ as
the corresponding functions $\bm{\varphi}({\bf x}) \in \Lambda$.

}

\vspace{0.2cm}

 Let us consider now a set of the functionals $I^{\nu}$, having
the form
\begin{equation}
\label{InuRhoPhi}
I^{\nu} \,\, = \,\, \int P^{\nu} \left( \bm{\rho}, \,
\bm{\rho}_{\bf x}, \, \bm{\phi}_{\bf x}, \, \bm{\rho}_{\bf xx},
\, \bm{\phi}_{\bf xx}, \dots \right) \,\, d^{d} x
\end{equation}

 Thus, we assume here that the functionals $I^{\nu}$ are
invariant with respect to the action of the group $G^{n_{2}}$
and the densities $P^{\nu}$ depend just on the derivatives of the
fields $\bm{\phi}$. The functionals (\ref{InuRhoPhi}) can be
considered on the space of rapidly decreasing functions just
putting
$$I^{\nu} \,\, = \,\,
\int_{-\infty}^{+\infty}\!\!\!\!\!\dots\int_{-\infty}^{+\infty}
P^{\nu} \left( \bm{\rho}, \,
\bm{\rho}_{\bf x}, \, \bm{\phi}_{\bf x}, \, \bm{\rho}_{\bf xx},
\, \bm{\phi}_{\bf xx}, \dots \right) \,\, d x^{1} \dots d x^{d} $$
or on the space of quasiperiodic functions, putting
$$I^{\nu} \,\, = \,\, \lim_{K\rightarrow\infty} \,\,
{1 \over (2 K)^{d}} \, \int_{-K}^{K}\!\!\dots\int_{-K}^{K}
P^{\nu} \left( \bm{\rho}, \,
\bm{\rho}_{\bf x}, \, \bm{\phi}_{\bf x}, \, \bm{\rho}_{\bf xx},
\, \bm{\phi}_{\bf xx}, \dots \right) \,\, d x^{1} \dots d x^{d} $$

 We will define also the variation derivatives of the
functionals $I^{\nu}$ using the variations of
$\bm{\rho}({\bf x})$, $\bm{\phi}({\bf x})$ with the same
(rapidly decreasing or quasiperiodic) properties as the original
functions. Easy to see then that in both cases just the standard 
Euler - Lagrange expressions for the variation derivatives can be used. 
It's not difficult to see also, that the functionals
$I^{\nu}$ are also well-defined on the functions from the
family $\Lambda$, having the form (\ref{RhoRepr}) - (\ref{PhiRepr}).

 Let us assume everywhere below that the functionals (\ref{InuRhoPhi})
are defined in the appropriate way in accordance with the corresponding
situation.

 The pairwise Poisson brackets of the densities
$\, P^{\nu} ({\bf x})$, $\, P^{\mu} ({\bf y})$ can be written in
the form:
$$\{ P^{\nu} ({\bf x}) \, , \, P^{\mu} ({\bf y}) \} \,\, = \,\,   
\sum_{l_{1},\dots,l_{d}} A^{\nu\mu}_{l_{1} \dots l_{d}}
(\bm{\rho}, \bm{\rho}_{\bf x}, \bm{\phi}_{\bf x}, \dots ) \,\,
\delta^{(l_{1})} (x^{1} - y^{1}) \, \dots \,
\delta^{(l_{d})} (x^{d} - y^{d}) $$
where
$$A^{\nu\mu}_{0 \dots 0}
(\bm{\rho}, \bm{\rho}_{\bf x}, \bm{\phi}_{\bf x}, \dots )
\,\,\, \equiv \,\,\, \partial_{x^{1}} \, Q^{\nu\mu 1}
(\bm{\rho}, \bm{\rho}_{\bf x}, \bm{\phi}_{\bf x}, \dots )
\, + \, \dots \, + \,
\partial_{x^{d}} \, Q^{\nu\mu d}
(\bm{\rho}, \bm{\rho}_{\bf x}, \bm{\phi}_{\bf x}, \dots ) $$  
for some functions
$Q^{\nu\mu q} (\bm{\rho}, \bm{\rho}_{\bf x}, \bm{\phi}_{\bf x},
\dots )$.

\vspace{0.2cm}

{\bf Definition 2.2.}

{\it We call a set $(I^{1}, \dots, I^{Q})$,
$\,\, Q =  m (d + 1) + n_{2} + s$,
of commuting functionals (\ref{InuRhoPhi}) a complete
Hamiltonian set on a regular Hamiltonian family $\Lambda$ of $m$-phase
solutions of system (\ref{rhophisystem}) with $n_{2}$ pseudo-phases
if:

1) The values of the functionals $(I^{1}, \dots, I^{Q})$
on any submanifold, given by the constraints
$${\bf p}_{1} \,\, = \,\, {\rm const} , \,\, \dots \, , \,\,
{\bf p}_{d} \,\, = \,\, {\rm const} \,\, ,$$
in the space of parameters on $\Lambda$,
give a complete set of parameters $(U^{1}, \dots, U^{Q})$
on this submanifold, excluding the initial phase shifts;

2) The Hamiltonian flows, generated by the functionals
$(I^{1}, \dots, I^{Q})$,
generate on $\Lambda$ linear phase shifts of $\bm{\theta}_{0}$
with frequencies $\bm{\omega}^{\nu} ({\bf U})$, and linear phase
shifts of $\bm{\tau}_{0}$ with frequencies
$\bm{\Omega}^{\nu} ({\bf U})$, such that
$${\rm rk} \,\,
\begin{Vmatrix}
\omega^{\alpha \nu} ({\bf U})  \\
\Omega^{j\nu} ({\bf U})
\end{Vmatrix}
\,\,\, = \,\,\, m \, + \, n_{2}$$

3) At every ``point'' of the submanifold $\Lambda$
the linear space generated by the variation derivatives
$\delta I^{\nu} / \delta \varphi^{i} ({\bf x})$ contains
the variation derivatives of all the corresponding annihilators 
of the bracket (\ref{MultDimPBr}), such that we can write
$$v^{(l)}_{i}({\bf x}, \,\, {\bf U}, \, \bm{\theta}_{0})
\,\, = \,\,
\sum_{\nu=1}^{Q} \gamma^{l}_{\nu} ({\bf U}) \,\, \left.
{\delta I^{\nu} \over \delta \varphi^{i} ({\bf x})} \right|_{\Lambda} $$
for some functions $\gamma^{l}_{\nu} ({\bf U})$ on the
family $\Lambda$.
}

\vspace{0.2cm}

 We can see then that in the presence of a complete Hamiltonian
set of the commuting functionals (\ref{InuRhoPhi}) the parameters
${\bf U}$ on the family $\Lambda$ can be also chosen in the form
$( U^{1}, \dots, U^{Q}, \, {\bf p}_{1}, \dots, {\bf p}_{d})$,
where $U^{\nu} = \langle P^{\nu} \rangle$. We will also assume here
that the Jacobian of the coordinate transformation
$$({\bf k}_{1}, \dots, {\bf k}_{d}, \, \bm{\omega},
{\bf p}_{1}, \dots, {\bf p}_{d}, \, \bm{\Omega},
n^{1}, \dots, n^{s}) \,\, \rightarrow \,\,
(U^{1}, \dots, U^{N}) $$
is different from zero whenever the values $U^{\nu}$ represent 
a complete set of parameters on $\Lambda$
excluding the initial phase shifts
$\, \bm{\theta}_{0}$, $\bm{\tau}_{0}$.

 Let us consider now the functionals
\begin{equation}
\label{Jnu}
J^{\nu} \,\, = \,\, \int_{0}^{2\pi}\!\!\!\!\!\dots\int_{0}^{2\pi}
P^{\nu} \left( \bm{\rho}, \,
k_{1}^{\beta_{1}} \bm{\rho}_{\theta^{\beta_{1}}},
\dots, k_{d}^{\beta_{d}} \bm{\rho}_{\theta^{\beta_{d}}}, \,
k_{1}^{\gamma_{1}} \bm{\phi}_{\theta^{\gamma_{1}}} + {\bf p}_{1},
\dots \right) \, {d^{m} \theta \over (2\pi)^{m}}
\end{equation}
on the space of $2\pi$-periodic in each $\theta^{\alpha}$ functions
$\bm{\rho}$ and $\bm{\phi}$.

 The variation derivatives of the functionals $J^{\nu}$
\begin{equation}
\label{VarDer}
\bm{\zeta}^{(\nu)}_{[{\bf U}]} (\bm{\theta} + \bm{\theta}_{0})
\,\, = \,\, \left(
\left. {\delta J^{\nu} \over \delta \rho^{1} (\bm{\theta})}
\right|_{\Lambda} , \dots ,
\left. {\delta J^{\nu} \over \delta \rho^{n_{1}} (\bm{\theta})}
\right|_{\Lambda} \, , \,
\left. {\delta J^{\nu} \over \delta \phi^{1} (\bm{\theta})}
\right|_{\Lambda}  , \dots ,
\left. {\delta J^{\nu} \over \delta \phi^{n_{2}} (\bm{\theta})}
\right|_{\Lambda}
\right)
\end{equation}
represent left eigen-vectors of the operator
${\hat L}^{i}_{j[{\bf U}, \bm{\theta}_{0}]}$ with zero
eigenvalues, which depend regularly on parameters ${\bf U}$
on $\Lambda$. Since the number of independent parameters
$({\bf k}_{1}, \dots, {\bf k}_{d})$ on $\Lambda$ is equal to
$m d$, we can claim that the number of linear independent vectors
(\ref{VarDer}) should not be less than $m +  n_{2} + s$ for a
complete Hamiltonian set of the functionals $I^{\nu}$ on $\Lambda$
according to the first requirement of Definition 2.2. Thus, we can
formulate here the following Proposition:

\vspace{0.2cm}

{\bf Proposition 2.1.}

{\it

Let the set of the functionals $(I^{1}, \dots, I^{Q})$
represent a complete Hamiltonian set on a regular Hamiltonian family
of $m$-phase solutions of (\ref{rhophisystem}) $\Lambda$ with
$n_{2}$ pseudo-phases. Then the linear span of the vectors
(\ref{VarDer}) contains all the regular left eigen-vectors
$\bm{\kappa}^{(q)}_{[{\bf U}]} (\bm{\theta} + \bm{\theta}_{0})$
of the operator ${\hat L}^{i}_{j[{\bf U}, \bm{\theta}_{0}]}$
with zero eigen-values.
}

\vspace{0.2cm}

 From the other hand, for a complete Hamiltonian family $\Lambda$
we can then claim also, that the number of the linearly independent
vectors (\ref{VarDer}) is exactly equal to the number of
$\bm{\kappa}^{(q)}_{[{\bf U}]} (\bm{\theta} + \bm{\theta}_{0})$
representing all the linearly independent regular left eigen-vectors
of the operator ${\hat L}^{i}_{j[{\bf U}, \bm{\theta}_{0}]}$
with zero eigen-values. As a corollary, we can formulate here
the following Lemma, which will be rather important in our further
considerations.

\vspace{0.2cm}

{\bf Lemma 2.1.}

{\it

 Let the set of the functionals $(I^{1}, \dots, I^{Q})$
represent a complete Hamiltonian set on a regular Hamiltonian family
$\Lambda$ of $m$-phase solutions of (\ref{rhophisystem}) with
$n_{2}$ pseudo-phases. Consider the corresponding functions
$$k^{\alpha}_{p} = k^{\alpha}_{p} \left(
U^{1}, \dots, U^{Q}, \, {\bf p}_{1}, \dots, {\bf p}_{d} \right)$$
in the coordinate system
$( U^{1}, \dots, U^{Q}, \, {\bf p}_{1}, \dots, {\bf p}_{d})$.
Then the functionals
$$k^{\alpha}_{p} \left(
J^{1}, \dots, J^{Q}, \, {\bf p}_{1}, \dots, {\bf p}_{d} \right)$$
have identically zero variation derivatives w.r.t. $\bm{\rho}$
and $\bm{\phi}$ on $\Lambda$.
}

\vspace{0.2cm}

 Proof.

 Indeed, the conditions of the Lemma imply that the number of the
linearly independent vectors 
$\bm{\zeta}^{(\nu)}_{[{\bf U}]} (\bm{\theta})$ on $\Lambda$
is equal to $\, m + n_{2} + s$. As a corollary, we can write
$m d$ independent relations
$$\sum_{\nu=1}^{Q} \lambda^{\tau}_{\nu}({\bf U}) \,\,
\bm{\zeta}^{(\nu)}_{[{\bf U}]} (\bm{\theta} + \bm{\theta}_{0})
\,\, \equiv \,\, 0 \,\,\,\,\, , \,\,\,\,\, \tau = 1, \dots, m d $$
with some functions $\lambda^{\tau}_{\nu}({\bf U})$ on $\Lambda$.

 For the corresponding coordinates $U^{\nu}$ on $\Lambda$ we can
write then the relations
$$\sum_{\nu=1}^{Q} \lambda^{\tau}_{\nu}({\bf U}) \, dU^{\nu}
\,\, = \,\, \sum_{q=1}^{d} \sum_{\beta=1}^{m}
\mu^{(\tau)}_{(\beta q)}({\bf U}) \, dk^{\beta}_{q}({\bf U}) $$
with some matrix $\mu^{(\tau)}_{(\beta q)}({\bf U})$.

 Since the values
$\, {\bf U} \, = \,
( U^{1}, \dots, U^{Q}, \, {\bf p}_{1}, \dots, {\bf p}_{d})$
represent a coordinate system on $\Lambda$, the matrix
$\mu^{(\tau)}_{(\beta q)}({\bf U})$ is invertible and we can
write the relations
$$d k^{\beta}_{q} \,\, = \,\, \sum_{\tau=1}^{md}
({\hat \mu}^{-1})^{(\beta q)}_{(\tau)}({\bf U}) \,
\sum_{\nu=1}^{N} \lambda^{(\tau)}_{\nu}({\bf U}) \,
dU^{\nu}$$
for every $k^{\beta}_{q}$, which gives the proof of the Lemma.

{\hfill Lemma 2.1 is proved.}

\vspace{0.2cm}

 As a corollary of Lemma 2.1 we can claim that the functionals  
$k^{\alpha}_{p} (I^{1}, \dots, I^{Q}, \, {\bf p}_{1}, \dots,
{\bf p}_{d})$ generate zero flows on the family $\Lambda$.
Using Definition 2.2 we can write then
\begin{equation}
\label{komegazeroflow}
\sum_{\nu=1}^{Q} \,\,
{\partial k^{\alpha}_{p} (U^{1}, \dots, U^{Q},
\, {\bf p}_{1}, \dots, {\bf p}_{d}) \over \partial U^{\nu}}
\,\,\,
\omega^{\beta \nu} (U^{1}, \dots, U^{Q}, \, {\bf p}_{1}, \dots,
{\bf p}_{d}) \,\,\, \equiv \,\,\, 0
\end{equation}
\begin{equation}
\label{kOMEGAzeroflow}
\sum_{\nu=1}^{Q} \,\,
{\partial k^{\alpha}_{p} (U^{1}, \dots, U^{Q},
\, {\bf p}_{1}, \dots, {\bf p}_{d}) \over \partial U^{\nu}}
\,\,\,
\Omega^{j \nu} (U^{1}, \dots, U^{Q}, \, {\bf p}_{1}, \dots,
{\bf p}_{d}) \,\,\, \equiv \,\,\, 0
\end{equation}
for the functions
$k^{\alpha}_{p} (U^{1}, \dots, U^{Q}, \, {\bf p}_{1},
\dots, {\bf p}_{d})$ on $\Lambda$.

 Finally, let us note also, that in the presence of a complete
Hamiltonian set $(I^{1}, \dots, I^{Q})$ for a regular Hamiltonian
family $\Lambda$ of $m$-phase solutions of (\ref{rhophisystem}) 
with $n_{2}$ pseudo-phases we can claim in fact, that the number 
of annihilators of the bracket (\ref{MultDimPBr}) on $\Lambda$ is
equal to the number of the additional parameters
$(n^{1}, \dots, n^{s})$. Indeed, according to the requirements
(2)-(3) of Definition 2.2, the number of the linearly independent
vectors (\ref{VarDer}) is equal to $m + n_{2} + s$, where $s$ is
the number of annihilators of the bracket (\ref{MultDimPBr}) on
$\Lambda$. Comparing this number with the number of the vectors
$\bm{\kappa}^{(q)}_{[{\bf U}]} (\bm{\theta} + \bm{\theta}_{0})$
we get the required statement.

\vspace{0.2cm}

 Let us discuss now the procedure of the bracket averaging.
Our considerations here will follow in many features the scheme
of \cite{DNMultDim,JMPMultDim}.

 Let us introduce the extended field space
$\bm{\varphi} ({\bf x}) \, \rightarrow \,
\bm{\varphi} (\bm{\theta}, {\bf X})$, where all the functions
$\bm{\varphi} (\bm{\theta}, {\bf X})$ are $2\pi$-periodic in
each $\theta^{\alpha}$, and consider the Poisson bracket
\begin{equation}
\label{EpsExtBracket}
\{\varphi^{i}(\bm{\theta}, {\bf X}) \, , \,
\varphi^{j}(\bm{\theta}^{\prime}, {\bf Y})\}
\,\,\, = \,\,\, \sum_{l_{1},\dots,l_{d}}  \!
\epsilon^{l_{1} + \dots + l_{d}} \,
B^{ij}_{(l_{1},\dots,l_{d})}
(\bm{\varphi}, \, \epsilon\, \bm{\varphi}_{\bf X}, \dots)
\,\, \delta_{l_{1} X^{1} \dots l_{d} X^{d}}({\bf X} - {\bf Y})
\,\, \delta (\bm{\theta} - \bm{\theta}^{\prime})
\end{equation}
on the space of fields $\bm{\varphi} (\bm{\theta}, {\bf X})$.

 For convenience we will define here the delta-function
$\delta (\bm{\theta} - \bm{\theta}^{\prime})$
and its higher derivatives
$\delta_{\theta^{\alpha_{1}}\dots\theta^{\alpha_{s}}}
(\bm{\theta} - \bm{\theta}^{\prime})$
on the space of $2\pi$-periodic functions by the formula
$$\int_{0}^{2\pi}\!\!\!\dots\int_{0}^{2\pi}
\delta_{\theta^{\alpha_{1}}\dots\theta^{\alpha_{s}}}
(\bm{\theta} - \bm{\theta}^{\prime}) \,\,
\psi (\bm{\theta}^{\prime}) \,\,
{d^{m} \theta^{\prime} \over (2\pi)^{m}} \,\,\, \equiv \,\,\,
\psi_{\theta^{\alpha_{1}}\dots\theta^{\alpha_{s}}} (\bm{\theta}) $$

 Also we put here the rule
$$\delta \, S \,\,\, \equiv \,\,\,
\int_{0}^{2\pi}\!\!\!\dots\int_{0}^{2\pi}
{\delta \, S \over \delta \varphi^{i} (\bm{\theta})} \,\,
\delta \varphi^{i} (\bm{\theta}) \,\,
{d^{m} \theta \over (2\pi)^{m}} $$
in the definition of the corresponding variation derivatives.

 Consider now the submanifold ${\mathcal K}$ in the extended field
space defined by the following conditions:

1) For given functions
$\{{\bf S}({\bf X}), \, \bm{\Sigma}({\bf X}), \, {\bf U}({\bf X})\}$
the functions $\bm{\varphi} (\bm{\theta}, {\bf X}) \in {\mathcal K}$
are defined by the formulas
\begin{multline}
\label{SubManK}
\rho^{i} (\bm{\theta}, {\bf X})
\,\,\, = \,\,\, R^{i} \left(
{{\bf S} ({\bf X}) \over \epsilon} \, + \,
\bm{\theta}, \,\, {\bf U} ({\bf X})  \right)
\,\,\, , \,\,\,\,\,\,\,\, i = 1, \dots, n_{1}  \, , \\
\phi^{j} (\bm{\theta}, {\bf X})
\, = \,\, \Psi^{j} \left(
{{\bf S} ({\bf X}) \over \epsilon} \, + \,
\bm{\theta}, \,\, {\bf U} ({\bf X}) \right)
\,\, + \,\, {1 \over \epsilon} \, \Sigma^{j} ({\bf X})
\,\,\, , \,\,\,\,\,\,\,\, j = 1, \dots, n_{2}
\end{multline}
where the values ${\bf U}$ represent the full set of parameters on
$\Lambda$ excluding the initial phase shifts.

2) The functions ${\bf U}({\bf X})$ are connected with the functions
${\bf S}({\bf X})$ and $\bm{\Sigma}({\bf X})$ by the relations
\begin{equation}
\label{kprelations}
k^{\alpha}_{q} \left( {\bf U}({\bf X}) \right) \,\, = \,\,
S^{\alpha}_{X^{q}} \,\,\,\,\, , \,\,\,\,\,
p^{j}_{q} \left( {\bf U}({\bf X}) \right) \,\, = \,\,
\Sigma^{j}_{X^{q}}
\end{equation}
where ${\bf k}_{q}$ and ${\bf p}_{q}$ are the corresponding wave
numbers and ``pseudo wave numbers'' defined on the family $\Lambda$.

 Thus, the elements
$\bm{\varphi} (\bm{\theta}, {\bf X}) \in {\mathcal K}$
are parametrized by the functions 
$\{{\bf S}({\bf X}), \, \bm{\Sigma}({\bf X}), \, {\bf U}({\bf X})\}$
with relations (\ref{kprelations})
and are connected with the zero approximation (\ref{rho0phi0}) for
the modulated $m$-phase solutions of (\ref{rhophisystem}). In the
presence of a complete Hamiltonian set of integrals $I^{\nu}$ the
parameters ${\bf U}$ can be chosen in the form
$(U^{1}, \dots, U^{Q}, \, {\bf p}_{1}, \dots, {\bf p}_{d})$,
where $U^{\nu} \equiv \langle P^{\nu} \rangle$ and ${\bf p}_{q}$
are given by relations (\ref{kprelations}).

 Let us introduce the functionals
\begin{equation}
\label{DefFuncSigma}
\Sigma^{i} ({\bf X}) \,\, = \,\, \epsilon \,
\int_{0}^{2\pi}\!\!\!\dots\int_{0}^{2\pi}
\, \phi^{j} (\bm{\theta}, {\bf X}) \,\,\,
{d^{m} \theta \over (2\pi)^{m}}
\end{equation}

 It is easy to see that the values of $\Sigma^{i} ({\bf X})$ on the
functions $\bm{\varphi} (\bm{\theta}, {\bf X}) \in {\mathcal K}$
coincide with the corresponding parameters on ${\mathcal K}$. We can
consider then the parameters $\bm{\Sigma} ({\bf X})$ and
$({\bf p}_{1}({\bf X}), \dots, {\bf p}_{d}({\bf X}))$ as the
functionals on the whole extended field space, having the
appropriate values on the submanifold ${\mathcal K}$.

 To introduce the analogous functionals for the parameters
$U^{\nu} ({\bf X})$ let us introduce the functionals
$$J^{\nu} ({\bf X}) \,\, = \,\,
\int_{0}^{2\pi}\!\!\!\dots\int_{0}^{2\pi}
P^{\nu} \left( \bm{\rho}, \, \epsilon \bm{\rho}_{\bf X}, \,
\epsilon \bm{\phi}_{\bf X}, \, \dots \right) \,
{d^{m} \theta \over (2\pi)^{m}}
\,\,\,\,\,\,\,\, , \,\,\,\,\,\,\,\,\,\, \nu = 1, \dots, Q , $$
and consider their values on the submanifold ${\mathcal K}$.

 Easy to see that we can write on ${\mathcal K}$:
\begin{equation}
\label{JUtransform}
J^{\nu}({\bf X}) \,\, = \,\, U^{\nu}({\bf X}) \, + \, \sum_{l\geq1}
\epsilon^{l} \, J^{\nu}_{(l)}({\bf X}) \,\,\,\,\,\,\,\, , \,\,\,\,\,
\nu = 1, \dots, Q
\end{equation}
where $J^{\nu}_{(l)}$ are some local functions of
$(U^{1} ({\bf X}), \dots, U^{Q}({\bf X}), \,
{\bf p}_{1} ({\bf X}), \dots, {\bf p}_{d} ({\bf X}))$
and their spatial derivatives which are polynomial in the derivatives
and have grading degree $l$ in terms of the total
number of differentiations with respect to ${\bf X}$.

 Let us say that the higher terms in (\ref{JUtransform}) are in 
fact not uniquely defined on ${\mathcal K}$ due to the compatibility 
relations (\ref{SpatialConstraints}). It is in fact sufficient for us 
that the terms $J^{\nu}_{(l)}({\bf X})$ can be chosen in some definite 
way in every order $l \geq 1$. Let us note also that the corresponding 
choice affects the definition of the functionals $U^{\nu} ({\bf X})$ 
just in the higher orders in $\epsilon$ ($l \geq 1$) which is actually 
not important for the construction.

 Transformation (\ref{JUtransform}) can be also inverted as
a formal series in $\epsilon$, such that we have
\begin{equation}
\label{UJtransform}
U^{\nu}({\bf X}) \,\, = \,\, J^{\nu}({\bf X}) \, + \, \sum_{l\geq1}
\epsilon^{l} \, U^{\nu}_{(l)}({\bf X}) \,\,\,\,\,\,\,\, , \,\,\,\,\,
\nu = 1, \dots, Q
\end{equation}
on the submanifold ${\mathcal K}$. Now the functions 
$U^{\nu}_{(l)}$ represent local functions of \linebreak
$(J^{1} ({\bf X}), \dots, J^{Q} ({\bf X}), \,
{\bf p}_{1} ({\bf X}), \dots, {\bf p}_{d} ({\bf X}))$
and their spatial derivatives, 
polynomial in the derivatives, and having degree $l$ in terms 
of the total number of differentiations w.r.t.
${\bf X}$. Now, we can consider the values $U^{\nu} ({\bf X})$
as the functionals on the whole extended field space.

 Let us put for simplicity the boundary conditions
$k^{\alpha}_{1} (X^{1}, 0, \dots, 0) \rightarrow 0$,
$X^{1} \rightarrow - \infty$, for the functionals
$k^{\alpha}_{1} ({\bf U}, \bm{\Sigma}_{\bf X})$ on the extended
functional space and define also the functionals
$S^{\alpha} ({\bf X})$ by the formula
\begin{equation}
\label{DefinitionS}
S^{\alpha}({\bf X})  =  \int_{-\infty}^{X^{1}} \!\!
k^{\alpha}_{1} (X^{\prime 1}, 0, \dots, 0) \, d X^{\prime 1}
\,\, + \, \dots \, +
\int_{0}^{X^{d}} \!\!\! k^{\alpha}_{d}
(X^{1}, \dots, X^{d-1}, X^{\prime d}) \, d X^{\prime d}
\end{equation}

 Now, all the parameters on the submanifold ${\mathcal K}$ are defined
as functionals on the whole extended field space. Let us note that
on the submanifold ${\mathcal K}$ we naturally have the relations
$S^{\alpha}_{X^{q}} = k^{\alpha}_{q} ({\bf X})$ which are in
general not true outside ${\mathcal K}$.

 Let us consider now the Poisson brackets of the functionals,
introduced above, on the submanifold ${\mathcal K}$. According to the
definition of the functionals $\bm{\Sigma} ({\bf X})$ and
Definition 2.2 it is not difficult to get the following relations
for the brackets of $\bm{\Sigma} ({\bf X})$ and
$U^{\mu}({\bf Y})$ on ${\mathcal K}$:
\begin{equation}
\label{SigmajSigmal}
\left. \left\{ \Sigma^{j} ({\bf X}) \, , \, \Sigma^{l} ({\bf Y})
\right\} \right|_{\mathcal K} \,\,\, = \,\,\, O (\epsilon^{2})
\,\,\,\,\, ,
\end{equation}
\begin{equation}
\label{SigmajUmu}
\left. \left\{ \Sigma^{j} ({\bf X}) \, , \, U^{\mu}({\bf Y})
\right\} \right|_{\mathcal K} \,\,\, = \,\,\, \epsilon \,
\Omega^{j \mu} ({\bf X}) \,\, \delta ({\bf X} - {\bf Y})
\,\,\, + \,\,\, O (\epsilon^{2}) \,\,\,\,\, ,
\end{equation}
$j, l = 1, \dots, n_{2}$, $\, \nu = 1, \dots, Q$.

 Using relations (\ref{kOMEGAzeroflow}) and
(\ref{SigmajSigmal}) - (\ref{SigmajUmu}) we can then write
\begin{equation}
\label{Sigmajkalphap}
\left. \left\{ \Sigma^{j} ({\bf X}) \, , \,
k^{\alpha}_{p} ({\bf Y}) \right\} \right|_{\mathcal K}
\,\,\, = \,\,\, O (\epsilon^{2})
\end{equation}
for the functionals
$k^{\alpha}_{p} ( U^{1}({\bf X}), \dots, U^{Q}({\bf X}),
\, \bm{\Sigma}_{X^{1}}, \dots, \bm{\Sigma}_{X^{d}} )$.

 The pairwise Poisson brackets of the functionals
$U^{\nu}({\bf X})$ have the order $O (\epsilon)$ everywhere
on the extended field space and we can write on ${\mathcal K}$:
$$\left. \left\{ U^{\nu} ({\bf X})\, , \,
U^{\mu} ({\bf Y}) \right\}  \right|_{\mathcal K} \,\,\, = \,\,\,
\left. \left\{ J^{\nu} ({\bf X})\, , \,
J^{\mu} ({\bf Y}) \right\} \right|_{\mathcal K}
\,\,\, + \,\,\, O (\epsilon^{2}) \,\,\, =  $$
$$= \,\,\, \epsilon \, \langle A^{\nu\mu}_{10\dots0} \rangle
({\bf X}) \,\,  \delta_{X^{1}} ({\bf X} - {\bf Y})
+ \,\, \dots \,\, + \,\,\,
\epsilon \, \langle A^{\nu\mu}_{0\dots01} \rangle
({\bf X}) \,\,\, \delta_{X^{d}} ({\bf X} - {\bf Y}) \,\,\, +  $$
$$+ \,\,\, \epsilon \, \left[ \langle Q^{\nu\mu \, p} \rangle
\right]_{X^{p}} \,\,\, \delta ({\bf X} - {\bf Y})
\,\,\, + \,\,\, O (\epsilon^{2})
\,\,\,\,\,\,\,\, , \,\,\,\,\,\,\,\,\,\,
\nu, \mu = 1, \dots, Q $$

 Let us prove here the following important Lemma:

\vspace{0.2cm}

{\bf Lemma 2.2.}

{\it
Let $(I^{1}, \dots, I^{Q})$, $Q = m (d + 1) + n_{2} + s$,
represent a complete Hamiltonian set of commuting functionals on
a regular Hamiltonian family $\Lambda$ of $m$-phase solutions of
(\ref{rhophisystem}) with $n_{2}$ pseudo-phases.
Consider the corresponding functions
$k^{\alpha}_{p} ({\bf U}) \, = \,
k^{\alpha}_{p} (U^{1}, \dots, U^{Q}, \,
{\bf p}_{1}, \dots, {\bf p}_{d})$ on the family $\Lambda$.

 Consider any functional ${\tilde I}$ of the form
\begin{equation}
\label{ItildeDefinition}
{\tilde I} \,\, = \,\, \int {\tilde P} \left( \bm{\rho}, \,
\bm{\rho}_{\bf x}, \, \bm{\phi}_{\bf x}, \, \bm{\rho}_{\bf xx},
\, \bm{\phi}_{\bf xx}, \dots \right) \,\, d^{d} x \,\,\, ,
\end{equation}
leaving the family $\Lambda$ and the parameters ${\bf U}$
invariant and generating on $\Lambda$
linear shifts of $\theta^{\alpha}_{0}$ with frequencies
${\tilde \omega}^{\alpha} ({\bf U})$ and linear shifts of
$\tau^{j}_{0}$ with frequencies ${\tilde \Omega}^{j} ({\bf U})$.
Let us consider the functionals
$${\tilde J} ({\bf X}) \,\, = \,\,
\int_{0}^{2\pi}\!\!\!\dots\int_{0}^{2\pi} {\tilde P}
\left( \bm{\rho}, \, \epsilon \bm{\rho}_{\bf X}, \,
\epsilon \bm{\phi}_{\bf X}, \, \epsilon^{2} \bm{\rho}_{\bf XX},
\, \epsilon^{2} \bm{\phi}_{\bf XX}, \dots \right) \,
{d^{m} \theta \over (2\pi)^{m}} $$

Then the functionals
$k^{\alpha}_{p} ( U^{1}({\bf X}), \dots, U^{Q}({\bf X}),
\, \bm{\Sigma}_{X^{1}}, \dots, \bm{\Sigma}_{X^{d}} ) $
have the following Poisson brackets with the functionals
${\tilde J} ({\bf Y})$ on ${\mathcal K}$:
$$\left. \left\{ k^{\alpha}_{p} ({\bf X})\, , \,
{\tilde J} ({\bf Y}) \right\} \right|_{\mathcal K} \,\,\, = \,\,\,
\epsilon \, \left[ \, {\tilde \omega}^{\alpha} ({\bf X}) \,\,
\delta ({\bf X} - {\bf Y}) \, \right]_{X^{p}}
\,\,\, + \,\,\, O (\epsilon^{2}) $$

}

Proof.

 Consider the dynamical system generated by the functional
\begin{equation}
\label{Jtildeq}
{\tilde J}_{[q]} \,\, = \,\, \int {\tilde J} ({\bf Y})
\,\, q ({\bf Y}) \,\, d^{d} Y
\end{equation}
with compactly supported $q ({\bf Y})$ according to bracket
(\ref{EpsExtBracket}).

 It is easy to see that in the main order $(O(1))$ the
corresponding evolution leaves invariant the submanifold
${\mathcal K}$, generating the shifts of the functions
${\bf S} ({\bf X})$ and $\bm{\Sigma} ({\bf X})$
with the frequencies
$\epsilon \, q ({\bf X}) \, {\tilde{\bm{\omega}}} ({\bf X})$ and
$\epsilon \, q ({\bf X}) \, {\tilde{\bm{\Omega}}} ({\bf X})$
respectively. As a result, we can decompose the dynamical system
on ${\mathcal K}$ into two parts:

1) The dynamics along the submanifold ${\mathcal K}$
giving the shifts of parameters
${\bf S} ({\bf X})$ and $\bm{\Sigma} ({\bf X})$
with the frequencies
$\epsilon \, q ({\bf X}) \, {\tilde{\bm{\omega}}} ({\bf X})$ and
$\epsilon \, q ({\bf X}) \, {\tilde{\bm{\Omega}}} ({\bf X})$;

2) The additional dynamics of the order $O (\epsilon)$ having
the form
$$\varphi^{i}_{t} \,\,\, = \,\,\, \epsilon \,
{\tilde \eta}^{i}_{[q]} \left(
{{\bf S} ({\bf X}) \over \epsilon} + \bm{\theta}, \, {\bf X}
\right) $$
with some $2\pi$-periodic in each $\theta^{\alpha}$ functions
${\tilde \eta}^{i}_{[q]} (\bm{\theta}, \, {\bf X} )$
on ${\mathcal K}$.

 The first part gives the following evolution of
$k^{\alpha}_{p} ({\bf X})$ on ${\mathcal K}$:
$$k^{\alpha}_{p \, t} \,\,\, = \,\,\, \epsilon \, \left(
q ({\bf X}) \,\, {\tilde \omega}^{\alpha} ({\bf X})
\right)_{X^{p}}$$
according to the definition of the functionals
$k^{\alpha}_{p} ({\bf X})$ on ${\mathcal K}$.

 To get the contribution of the second part to the evolution of
$k^{\alpha}_{p} ({\bf X})$ we can change in the main part the
functionals
$k^{\alpha}_{p} ( U^{1}({\bf X}), \dots, U^{Q}({\bf X}), \,
\bm{\Sigma}_{X^{1}}, \dots, \bm{\Sigma}_{X^{d}})$ to 
$k^{\alpha}_{p} ( J^{1}({\bf X}), \dots, J^{Q}({\bf X}), \,
\bm{\Sigma}_{X^{1}}, \dots, \bm{\Sigma}_{X^{d}})$ using
(\ref{JUtransform}) - (\ref{UJtransform}).
It's not difficult to see then that the main contribution of
the corresponding dynamics to the evolution of
$k^{\alpha}_{p} ({\bf X})$ is given by the convolution of the
variation derivatives of the corresponding functionals
$k^{\alpha}_{p} ( J^{1}, \dots, J^{Q}, \,
{\bf p}_{1}, \dots, {\bf p}_{d})$,
defined on the space of $2\pi$-periodic in each $\theta^{\alpha}$
functions $(\bm{\rho}(\bm{\theta}), \, \bm{\phi}(\bm{\theta}))$,
with the functions
${\tilde \eta}^{i}_{[q]} (\bm{\theta}, \, {\bf X} )$
at every given ${\bf X}$. According to Lemma 2.1 we get then that
the corresponding contribution is absent in the order
$O (\epsilon)$.

 Finally, we can write on ${\mathcal K}$:
$$\left. \left\{ k^{\alpha}_{p} ({\bf X})\, , \,
{\tilde J}_{[q]} \right\} \right|_{\mathcal K} \,\,\, = \,\,\,
\epsilon \, \left( q ({\bf X}) \,\,
{\tilde \omega}^{\alpha} ({\bf X}) \right)_{X^{p}}$$
which is equivalent to the assertion of the Lemma.

{\hfill Lemma 2.2 is proved.}

\vspace{0.2cm}

 As a corollary from Lemma 2.2 we can write, in particular
\begin{equation}
\label{kalphapUmu}
\left. \left\{ k^{\alpha}_{p} ({\bf X})\, , \,
U^{\mu}({\bf Y}) \right\} \right|_{\mathcal K} \,\,\, = \,\,\,
\epsilon \, \left[ \, \omega^{\alpha \mu} ({\bf X}) \,\,
\delta ({\bf X} - {\bf Y}) \, \right]_{X^{p}}
\,\,\, + \,\,\, O (\epsilon^{2})
\end{equation}
for the functionals $U^{\mu}({\bf Y})$, using the analogous
relations for $J^{\mu}({\bf Y})$ and relations
(\ref{UJtransform}).

 From relations (\ref{komegazeroflow}), (\ref{Sigmajkalphap})
and (\ref{kalphapUmu}) it is not difficult to get then also the
following relations on ${\mathcal K}$:
\begin{equation}
\label{kalphapkbetaq}
\left. \left\{ k^{\alpha}_{p} ({\bf X})\, , \,
k^{\beta}_{q} ({\bf Y}) \right\} \right|_{\mathcal K}
\,\,\, = \,\,\, O (\epsilon^{2})
\end{equation}

 Using the definition (\ref{DefinitionS}) of the functionals
$S^{\alpha} ({\bf X})$ and relations (\ref{Sigmajkalphap}),
(\ref{kalphapUmu}), (\ref{kalphapkbetaq}), we can write then
the following relations for their Poisson brackets on ${\mathcal K}$:
\begin{equation}
\label{SBrackets}
\begin{array}{c}
\left. \left\{ S^{\alpha} ({\bf X})\, , \, \Sigma^{j} ({\bf Y})
\right\} \right|_{\mathcal K} \,\,\, = \,\,\, O (\epsilon^{2})
\,\,\,\,\, , \,\,\,\,\,
\left. \left\{ S^{\alpha} ({\bf X})\, , \,
S^{\beta} ({\bf Y}) \right\} \right|_{\mathcal K}
\,\,\, = \,\,\, O (\epsilon^{2})  \\   \\
\left. \left\{ S^{\alpha} ({\bf X})\, , \,
U^{\mu}({\bf Y}) \right\} \right|_{\mathcal K} \,\,\, = \,\,\,
\epsilon \, \omega^{\alpha \mu} ({\bf X}) \,\,
\delta ({\bf X} - {\bf Y}) \,\,\, + \,\,\, O (\epsilon^{2})
\end{array}
\end{equation}

 It will be convenient now to choose the parameters
on the family $\Lambda$ in the form
\begin{equation}
\label{kpUcoordsystem}
\left( {\bf k}_{1}, \dots, {\bf k}_{d}, \,
{\bf p}_{1}, \dots, {\bf p}_{d}, \,
U^{1}, \dots, U^{m + n_{2} + s} \right)
\end{equation}
where $U^{\gamma} \equiv \langle P^{\gamma} \rangle$,
$\gamma = 1, \dots, m + n_{2} + s$, represent just a subset
of the set $U^{\nu}$,
$\nu = 1, \dots, Q = m (d + 1) + n_{2} + s$, and to consider the
functionals
$$\left\{ {\bf S} ({\bf X}), \, \bm{\Sigma} ({\bf X}), \,
U^{1} ({\bf X}), \dots, U^{m + n_{2} + s} ({\bf X}) \right\} $$
as completely independent ``coordinates'' on ${\mathcal K}$ according
to (\ref{kprelations}). Let us say that the subset
$\{ U^{\gamma} \}$ can be chosen in arbitrary way just to give
a functionally independent system (\ref{kpUcoordsystem}).
For convenience, we will denote now by ${\bf U}$ just a set
of the functionals $U^{\gamma}$:
${\bf U} = (U^{1}, \dots, U^{m + n_{2} + s})$.

 Let us introduce also the ``constraints''
$g^{i} (\bm{\theta}, {\bf X})$ near ${\mathcal K}$ just putting 
in general form:
$$g^{i} (\bm{\theta}, {\bf X}) \,\,\, = \,\,\,
\varphi^{i} (\bm{\theta}, {\bf X}) \,\, - \,\,
\Phi^{i} \left( {{\bf S}({\bf X}) \over \epsilon} + \bm{\theta},
\,\, {\bf S}_{\bf X}, \, \bm{\Sigma}_{\bf X}, \, {\bf U} ({\bf X}),
\, \bm{\Sigma} ({\bf X}) \right) $$
where $\Phi^{i}$ represent the right-hand part of relations
(\ref{SubManK}). We have to note that the functionals
$g^{i} (\bm{\theta}, {\bf X})$ are not independent. Thus, the
following relations for the ``gradients'' of
$g^{i} (\bm{\theta}, {\bf X})$ can be written on ${\mathcal K}$:
\begin{equation}
\label{ConstrDepend}
\int \int_{0}^{2\pi}\!\!\!\!\!\dots\int_{0}^{2\pi}
\left. {\delta G ({\bf Z}) \over
\delta \varphi^{i} (\bm{\theta}, {\bf X})}\right|_{\mathcal K}
\,\, \left. {\delta g^{i} (\bm{\theta}, {\bf X}) \over
\delta \varphi^{j} (\bm{\theta}^{\prime}, {\bf Y})}\right|_{\mathcal K}
\,\,\, {d^{m} \theta \over (2\pi)^{m}} \,\, d^{d} X
\,\,\, \equiv \,\,\, 0
\end{equation}
where $G ({\bf Z})$ represents any of the functionals
$S^{\alpha} ({\bf Z})$, $\bm{\Sigma}^{j} ({\bf Z})$ or
$U^{\gamma} ({\bf Z})$.

 Using Lemma 2.2 we can write also the relations
$$\left. \left\{ g^{i} (\bm{\theta}, {\bf X}) \, , \,
{\tilde J}_{[q]} \right\} \right|_{\mathcal K} \,\,\, = \,\,\,
O (\epsilon) $$
for any functional ${\tilde J}_{[q]}$ defined by (\ref{Jtildeq})
with ${\tilde J} ({\bf Y})$ satisfying the requirements of
Lemma 2.2. In particular, for the functionals
$$J_{[{\bf q}]} \,\,\, = \,\,\, \int J^{\mu} ({\bf Y}) \,\,
q_{\mu} ({\bf Y}) \,\, d^{d} Y \,\,\, , \,\,\,\,\,
U_{[{\bf q}]} \,\,\, = \,\,\, \int U^{\mu} ({\bf Y}) \,\,
q_{\mu} ({\bf Y}) \,\, d^{d} Y $$
with compactly supported $q_{\mu} ({\bf Y})$,
$\mu = 1, \dots, Q$, we can write
\begin{equation}
\label{giJq}
\left. \left\{ g^{i} (\bm{\theta}, {\bf X}) \, , \,
J_{[{\bf q}]} \right\} \right|_{\mathcal K} \,\,\, = \,\,\,
O (\epsilon) \,\,\, , \,\,\,\,\,
\left. \left\{ g^{i} (\bm{\theta}, {\bf X}) \, , \,
U_{[{\bf q}]} \right\} \right|_{\mathcal K} \,\,\, = \,\,\,
O (\epsilon)
\end{equation}

 Using the definition of the functionals $\Sigma^{j} ({\bf X})$
and relations (\ref{SBrackets}) we can write the same
relations also for the functionals
$\, \Sigma_{[{\bf p}]} \, = \,
\int \Sigma^{j} ({\bf Y}) \,\, p_{j} ({\bf Y}) \,\, d^{d} Y $,
i.e.
\begin{equation}
\label{giSigmap}
\left. \left\{ g^{i} (\bm{\theta}, {\bf X}) \, , \,
\Sigma_{[{\bf p}]} \right\} \right|_{\mathcal K} \,\,\, = \,\,\,
O (\epsilon)
\end{equation}

 We will need now another important Lemma:

\vspace{0.2cm}

{\bf Lemma 2.3.}

{\it Let $\Lambda$ be a complete regular family of $m$-phase
solutions of system (\ref{rhophisystem}) with $n_{2}$
pseudo-phases and system
(\ref{ortcond}) - (\ref{SpatialConstraints})
represent the corresponding regular Whitham system on $\Lambda$.
Let system (\ref{rhophisystem}) has the first integral
${\tilde I}$ of the form (\ref{ItildeDefinition}) such that we
have
$${\tilde P}_{t} \left( \bm{\rho}, \, \bm{\rho}_{\bf x}, \,
\bm{\phi}_{\bf x}, \dots \right) \,\, = \,\,
{\tilde Q}^{1}_{x^{1}} \left( \bm{\rho}, \,
\bm{\rho}_{\bf x}, \, \bm{\phi}_{\bf x}, \dots \right) \, + \,
\dots \, + \, {\tilde Q}^{d}_{x^{d}} \left( \bm{\rho}, \,
\bm{\rho}_{\bf x}, \, \bm{\phi}_{\bf x}, \dots \right) $$
on the solutions of (\ref{rhophisystem}). Then the Whitham
system (\ref{ortcond}) - (\ref{SpatialConstraints}) implies
the relation
$$\langle {\tilde P} \rangle_{T} \,\, = \,\,
\langle {\tilde Q}^{1} \rangle_{X^{1}}
\, + \, \dots \, + \, \langle {\tilde Q}^{d} \rangle_{X^{d}}
\,\,\, . $$

}

\vspace{0.2cm}

 Proof.

 Easy to see that for any time dependence of the parameters
on the family $\Lambda$ we can write:
$$\langle {\tilde P} \rangle_{T} \,\,\, = \,\,\,
\int_{0}^{2\pi}\!\!\!\!\!\dots\int_{0}^{2\pi}
\left( {\delta {\tilde J} \over \delta \rho^{i} (\bm{\theta})}
\Big\vert_{\Lambda} R^{i}_{T} (\bm{\theta}) \,\,\, + \,\,\,
{\delta {\tilde J} \over \delta \phi^{j} (\bm{\theta})}
\Big\vert_{\Lambda} \Psi^{j}_{T} (\bm{\theta}) \right) \,
{d^{m} \theta \over (2\pi)^{m}} \,\,\, +   $$
$$+ \,\,\, k^{\beta}_{q \, T} \,\,
{\partial {\tilde J} \over
\partial k^{\beta}_{q}} \Big\vert_{\Lambda}
\,\,\, + \,\,\, p^{j}_{q \, T} \,\,
{\partial {\tilde J} \over
\partial p^{j}_{q}} \Big\vert_{\Lambda} $$
where the functional
$${\tilde J} \,\, = \,\,
\int_{0}^{2\pi}\!\!\!\!\!\dots\int_{0}^{2\pi}
{\tilde P} \left( \bm{\rho}, \,
k_{1}^{\beta_{1}} \bm{\rho}_{\theta^{\beta_{1}}},
\dots, k_{d}^{\beta_{d}} \bm{\rho}_{\theta^{\beta_{d}}}, \,
k_{1}^{\gamma_{1}} \bm{\phi}_{\theta^{\gamma_{1}}} + {\bf p}_{1},
\dots \right) \, {d^{m} \theta \over (2\pi)^{m}} $$
is defined on the space of $2\pi$-periodic functions for any
given parameters $({\bf k}_{1}, \dots, {\bf k}_{d})$,
$({\bf p}_{1}, \dots, {\bf p}_{d})$.

 Let us also introduce the functions
$${\tilde \Pi}^{(l_{1} \dots l_{d})}_{\bm{\rho} \, i}
(\bm{\rho}, \, \bm{\rho}_{\bf x}, \, \bm{\phi}_{\bf x}, \dots )
\,\,\, \equiv \,\,\,
{\partial {\tilde P} (\bm{\rho}, \, \bm{\rho}_{\bf x}, \,
\bm{\phi}_{\bf x}, \dots ) \over \partial
\rho^{i}_{l_{1} x^{1} \dots l_{d} x^{d}} } \,\,\,\,\, ,  $$
$${\tilde \Pi}^{(l_{1} \dots l_{d})}_{\bm{\phi} \, j}
(\bm{\rho}, \, \bm{\rho}_{\bf x}, \, \bm{\phi}_{\bf x}, \dots )
\,\,\, \equiv \,\,\,
{\partial {\tilde P} (\bm{\rho}, \, \bm{\rho}_{\bf x}, \,
\bm{\phi}_{\bf x}, \dots ) \over \partial
\phi^{j}_{l_{1} x^{1} \dots l_{d} x^{d}} }
\,\,\,\,\, , \,\,\,\,\,\,\,\, l_{1}, \dots l_{d} \, \geq \, 0$$

 We can then write according to (\ref{rhophisystem})
$$\epsilon \, {\tilde Q}^{1}_{X^{1}} \, + \, \dots \, + \,
\epsilon \, {\tilde Q}^{d}_{X^{d}} \,
\,\,\, \equiv  \,\,\, \sum_{l_{1}, \dots, l_{d}} 
\epsilon^{l_{1} + \dots + l_{d}} \left(
{\tilde \Pi}^{(l_{1} \dots l_{d})}_{\bm{\rho} \, i} \,
A^{i}_{l_{1} x^{1} \dots l_{d} x^{d}} \, + \,
{\tilde \Pi}^{(l_{1} \dots l_{d})}_{\bm{\phi} \, j} \,
B^{j}_{l_{1} x^{1} \dots l_{d} x^{d}} \right) $$

 The fulfillment of conditions (\ref{SpatialConstraints})
permits us to introduce the functions ${\bf S} ({\bf X})$,
$\, \bm{\Sigma} ({\bf X})$ and consider the functions
${\tilde P} (\bm{\rho}, \bm{\rho}_{\bf x},
\bm{\phi}_{\bf x}, \dots )$ and
${\tilde Q} (\bm{\rho}, \bm{\rho}_{\bf x},
\bm{\phi}_{\bf x}, \dots )$ on the submanifold ${\mathcal K}$.

 Easy to see that the operators
$\epsilon \, \partial / \partial X^{p}$ on the submanifold
${\mathcal K}$ can be naturally represented as a sum of
$k^{\alpha}_{p} \, \partial / \partial \theta^{\alpha}$
and the terms proportional to $\epsilon$. So, let us introduce
on ${\mathcal K}$ the natural expansion for any expression
$f (\bm{\rho}, \bm{\rho}_{\bf x}, \bm{\phi}_{\bf x}, \dots )$
invariant under the action of the pseudo-phase group:
$$f (\bm{\rho}, \bm{\rho}_{\bf x}, \bm{\phi}_{\bf x}, \dots )
\Big\vert_{\mathcal K} \,\, = \,\, \sum_{l\geq0} \epsilon^{l} \,
f_{[l]} \left[ {{\bf S} ({\bf X}) \over \epsilon} + \bm{\theta};
\,\, {\bf k}_{1}, \dots, {\bf k}_{d},
{\bf p}_{1}, \dots, {\bf p}_{d}, {\bf U} \right] $$
where $f_{[l]}$ are smooth functions of the arguments
$({\bf k}_{1}, \dots, {\bf k}_{d},
{\bf p}_{1}, \dots, {\bf p}_{d}, {\bf U})$ and their
${\bf X}$-derivatives, polynomial in the derivatives and having
degree $l$ in terms of the total number of differentiations of these
parameters w.r.t. ${\bf X}$. Since the common phase shift is not
important in the integration w.r.t. $\bm{\theta}$, let us also
assume below that the phase shift
${\bf S} ({\bf X}) / \epsilon$ is omitted in the functions
$f_{[l]}$ after taking all the differentiations w.r.t. ${\bf X}$.

 We can write then (summation over all the repeated indexes):
$$\langle {\tilde Q}^{1} \rangle_{X^{1}} \, + \, \dots \, + \,
\langle {\tilde Q}^{d} \rangle_{X^{d}} \,\, = \,\,
\int_{0}^{2\pi}\!\!\!\!\!\dots\int_{0}^{2\pi} \left(
{\tilde Q}^{1}_{X^{1} [1]} \, + \, \dots \, + \,
{\tilde Q}^{d}_{X^{d} [1]} \right)
\, {d^{m} \theta \over (2\pi)^{m}} \,\, = $$
$$= \,\, \int_{0}^{2\pi}\!\!\!\!\!\dots\int_{0}^{2\pi}
\sum_{l_{1}, \dots, l_{d}}
\left( {\tilde \Pi}^{(l_{1} \dots l_{d})}_{\bm{\rho} \, i \, [0]}
\,\, A^{i}_{l_{1} X^{1} \dots l_{d} X^{d} \, [1]} \,\,\, + \,\,\,
{\tilde \Pi}^{(l_{1} \dots l_{d})}_{\bm{\rho} \, i \, [1]} \,\,
A^{i}_{l_{1} X^{1} \dots l_{d} X^{d} \, [0]} \right. \,\,\, + $$
$$+ \,\,\, \left.
{\tilde \Pi}^{(l_{1} \dots l_{d})}_{\bm{\phi} \, j \, [0]} \,\,
B^{j}_{l_{1} X^{1} \dots l_{d} X^{d} \, [1]} \,\,\, + \,\,\,
{\tilde \Pi}^{(l_{1} \dots l_{d})}_{\bm{\phi} \, j \, [1]} \,\,
B^{j}_{l_{1} X^{1} \dots l_{d} X^{d} \, [0]} \right)
{d^{m} \theta \over (2\pi)^{m}} \,\,\, = $$
$$= \int_{0}^{2\pi}\!\!\!\!\!\dots\int_{0}^{2\pi}
\sum_{l_{1}, \dots, l_{d}} \left(
{\tilde \Pi}^{(l_{1} \dots l_{d})}_{\bm{\rho} \, i \, [0]} \,\,
k^{\gamma^{1}_{1}}_{1} \dots k^{\gamma^{1}_{l_{1}}}_{1} \, \dots \,
k^{\gamma^{d}_{1}}_{d} \dots k^{\gamma^{d}_{l_{d}}}_{d} \,\,\,
A^{i}_{[1] \, \theta^{\gamma^{1}_{1}}\dots\theta^{\gamma^{1}_{l_{1}}}
\, \dots \, \theta^{\gamma^{d}_{1}}\dots\theta^{\gamma^{d}_{l_{d}}}}
\,\, +  \right. $$
$$+ \,\,\,
{\tilde \Pi}^{(l_{1} \dots l_{d})}_{\bm{\phi} \, j \, [0]} \,\,
k^{\gamma^{1}_{1}}_{1} \dots k^{\gamma^{1}_{l_{1}}}_{1} \,\, \dots \,\,
k^{\gamma^{d}_{1}}_{d} \dots k^{\gamma^{d}_{l_{d}}}_{d} \,\,\,
B^{j}_{[1] \, \theta^{\gamma^{1}_{1}}\dots\theta^{\gamma^{1}_{l_{1}}}
\, \dots \, \theta^{\gamma^{d}_{1}}\dots\theta^{\gamma^{d}_{l_{d}}}}
\,\,\,\,\, +  $$
$$+ \,\,\,\,\,
{\tilde \Pi}^{(l_{1} \dots l_{d})}_{\bm{\rho} \, i \, [0]} \,\,
\left( \omega^{\beta}  R^{i}_{\theta^{\beta}}
\right)_{l_{1} X^{1} \dots l_{d} X^{d} \, [1]} \,\,\, + \,\,\,
{\tilde \Pi}^{(l_{1} \dots l_{d})}_{\bm{\rho} \, i \, [1]} \,
\omega^{\beta} \,
R^{i}_{\theta^{\beta} l_{1} X^{1} \dots l_{d} X^{d} \, [0]}
\,\,\,\,\, +$$
$$+ \, \left.
{\tilde \Pi}^{(l_{1} \dots l_{d})}_{\bm{\phi} \, j \, [0]} \,\,
\left( \Omega^{j} +
\omega^{\beta} \Psi^{j}_{\theta^{\beta}}
\right)_{l_{1} X^{1} \dots l_{d} X^{d} \, [1]} \,\, + \,\,
{\tilde \Pi}^{(l_{1} \dots l_{d})}_{\bm{\phi} \, j \, [1]} \,
\omega^{\beta} \,
\Psi^{j}_{\theta^{\beta} l_{1} X^{1} \dots l_{d} X^{d} \, [0]}
\right) {d^{m} \theta \over (2\pi)^{m}} \, = $$
$$= \,\,\, \int_{0}^{2\pi}\!\!\!\!\!\dots\int_{0}^{2\pi}
\left( {\delta {\tilde J} \over \delta \rho^{i} (\bm{\theta})}
\Big\vert_{\Lambda} \,\, A^{i}_{[1]} (\bm{\theta}, {\bf X})
\,\,\, + \,\,\,
{\delta {\tilde J} \over \delta \phi^{j} (\bm{\theta})}
\Big\vert_{\Lambda} \,\, B^{j}_{[1]} (\bm{\theta}, {\bf X})
\right. \,\,\, + $$
$$+ \sum_{l_{1}, \dots, l_{d}} \left( \omega^{\beta}_{X^{1}} \,
{\tilde \Pi}^{(l_{1} \dots l_{d})}_{\bm{\rho} \, i \,[0]} \, l_{1}
R^{i}_{\theta^{\beta} \, (l_{1}-1) X^{1} \dots l_{d} X^{d} \, [0]}
\,\, + \,\, \omega^{\beta}_{X^{1}} \,
{\tilde \Pi}^{(l_{1} \dots l_{d})}_{\bm{\phi} \, j \,[0]} \, l_{1}
\Psi^{j}_{\theta^{\beta} \, (l_{1}-1) X^{1} \dots l_{d} X^{d} \, [0]}
\, +  \right. $$
$$\left. \dots  \, + \, \omega^{\beta}_{X^{d}} \,
{\tilde \Pi}^{(l_{1} \dots l_{d})}_{\bm{\rho} \, i \,[0]} \, l_{d}
R^{i}_{\theta^{\beta} \, l_{1} X^{1} \dots (l_{d} - 1) X^{d} \, [0]}
\, + \, \omega^{\beta}_{X^{d}} \,
{\tilde \Pi}^{(l_{1} \dots l_{d})}_{\bm{\phi} \, j \,[0]} \, l_{d}
\Psi^{j}_{\theta^{\beta} \,
l_{1} X^{1} \dots (l_{d} - 1) X^{d} \, [0]} \right) + $$
$$+ \,\,\, \Omega^{j}_{X^{1}}
{\tilde \Pi}^{(1 0 \dots 0)}_{\bm{\phi} \, j \,[0]}
\,\, + \,\, \dots \,\, + \,\, \Omega^{j}_{X^{d}}
{\tilde \Pi}^{(0 \dots 0 1)}_{\bm{\phi} \, j \,[0]} \,\,\, +$$
$$+ \,\,\, \sum_{l_{1}, \dots, l_{d}} \left(
\omega^{\beta} \,
{\tilde \Pi}^{(l_{1} \dots l_{d})}_{\bm{\rho} \, i \,[0]} \,\,
R^{i}_{\theta^{\beta} l_{1} X^{1} \dots l_{d} X^{d} \, [1]}
\,\, + \,\, \omega^{\beta} \,
{\tilde \Pi}^{(l_{1} \dots l_{d})}_{\bm{\rho} \, i \,[1]} \,\,
R^{i}_{\theta^{\beta} l_{1} X^{1} \dots l_{d} X^{d} \, [0]}
\,\,\, + \right. $$
$$\left. \left. + \,\,\, \omega^{\beta} \,
{\tilde \Pi}^{(l_{1} \dots l_{d})}_{\bm{\phi} \, j \,[0]} \,\,
\Psi^{j}_{\theta^{\beta} l_{1} X^{1} \dots l_{d} X^{d} \, [1]}
\,\, + \,\, \omega^{\beta} \,
{\tilde \Pi}^{(l_{1} \dots l_{d})}_{\bm{\phi} \, j \,[1]} \,\,
\Psi^{j}_{\theta^{\beta} l_{1} X^{1} \dots l_{d} X^{d} \, [0]}
\right) \right) \, {d^{m} \theta \over (2\pi)^{m}} $$

 We can see that the last four terms in the expression above
represent the integral of the value
$\, \omega^{\beta} \, \partial {\tilde P}_{[1]} /
\partial \theta^{\beta}$ and so are equal to zero. It's not difficult
to see now that the expression
$\, \langle {\tilde P} \rangle_{T} \, - \,
\langle {\tilde Q}^{1} \rangle_{X^{1}}
\, - \, \dots \, - \, \langle {\tilde Q}^{d} \rangle_{X^{d}} \, $
can be written in the form:
$$\int_{0}^{2\pi}\!\!\!\!\!\dots\int_{0}^{2\pi}
\left( {\delta {\tilde J} \over \delta \rho^{i} (\bm{\theta})}
\Big\vert_{\Lambda} \, \left( R^{i}_{T} \, - \, A^{i}_{[1]}
\right) \,\,\, + \,\,\,
{\delta {\tilde J} \over \delta \phi^{j} (\bm{\theta})}
\Big\vert_{\Lambda} \,
\left( \Psi^{j}_{T} \, - \, B^{j}_{[1]}
\right)  \right) {d^{m} \theta \over (2\pi)^{m}} \,\,\, +$$
$$+ \,\,\, \left( k^{\beta}_{q \, T} - \omega^{\beta}_{X^{q}}
\right) \, {\partial {\tilde J} \over \partial k^{\beta}_{q}} \,
\Big\vert_{\Lambda} \,\,\, + \,\,\,
\left( p^{j}_{q \, T} - \Omega^{j}_{X^{q}} \right) \,
{\partial {\tilde J} \over
\partial p^{j}_{q}} \, \Big\vert_{\Lambda} $$

 The last two terms of the above expression are obviously equal
to zero according to relations (\ref{TimeConstraints}). As for the
first term, we can see that it represents the inner product of
the variation derivative of the functional ${\tilde J}$ on
$\Lambda$ with the first $\epsilon$-discrepancy of system
(\ref{epsrhophisystem}) after the substitution of the main term
(\ref{rho0phi0}). Since the variation derivative of the functional
${\tilde J}$ on $\Lambda$ represents a regular left eigen-vector
of the corresponding linear operator
${\hat L}_{[{\bf k}_{1}, \dots, {\bf k}_{d},
{\bf p}_{1}, \dots, {\bf p}_{d}, {\bf U}]}$ with zero eigen-value,
we can claim that it is given by a linear combination of the
corresponding vectors
$\bm{\kappa}^{(q)}_{[{\bf k}_{1}, \dots, {\bf k}_{d},
{\bf p}_{1}, \dots, {\bf p}_{d}, {\bf U}]} (\bm{\theta})$
on the complete regular family $\Lambda$. We can see then that
the first term of the above expression is equal to zero view
relations (\ref{ortcond}).

{\hfill Lemma 2.3 is proved.}

\vspace{0.2cm}

 Using Lemma 2.3 we can replace in fact the Whitham system
(\ref{ortcond}) - (\ref{SpatialConstraints}) by the equivalent
system
\begin{equation}
\begin{array}{c}
\label{SSigmaUWhithSyst}
S^{\alpha}_{T} \,\, = \,\, \omega^{\alpha}
\left( {\bf S}_{\bf X}, \, \bm{\Sigma}_{\bf X}, \, {\bf U} \right)
\,\,\,\,\, , \,\,\,\,\,\,\,\,
\Sigma^{j}_{T} \,\, = \,\, \Omega^{j}
\left( {\bf S}_{\bf X}, \, \bm{\Sigma}_{\bf X}, \, {\bf U} \right)
\,\,\, , \\   \\
U^{\gamma}_{T} \,\, = \,\, \langle Q^{1 \gamma} \rangle_{X^{1}}
\,\, + \,\, \dots \,\, + \,\,
\langle Q^{d \gamma} \rangle_{X^{d}} \,\,\, ,
\end{array}
\end{equation}
$\gamma = 1, \dots, m + n_{2} + s$, for the coordinates
$$\left( {\bf S} ({\bf X}), \, \bm{\Sigma} ({\bf X}), \,
U^{1} ({\bf X}), \dots, U^{m + n_{2} + s} ({\bf X}) \right) $$
on the submanifold ${\mathcal K}$.

 Let us note also that in the case when the values
$$({\bf k}_{1}, \dots, {\bf k}_{d}, \,
{\bf p}_{1}, \dots, {\bf p}_{d}, \,
U^{1}, \dots, U^{m + n_{2} + s})$$
give the full set of parameters
on $\Lambda$ (excluding the initial phase shifts), it is not
difficult to get from Proposition 2.1 and Lemma 2.1 that the
linear span of variation derivatives of the functionals
$J^{\gamma}$, $\gamma = 1, \dots, m + n_{2} + s$, contain
all the regular left eigen-vectors
$\bm{\kappa}^{(q)}_{[{\bf k}_{1}, \dots, {\bf k}_{d},
{\bf p}_{1}, \dots, {\bf p}_{d}, {\bf U}]} (\bm{\theta})$
of the operator
${\hat L}_{[{\bf k}_{1}, \dots, {\bf k}_{d},
{\bf p}_{1}, \dots, {\bf p}_{d}, {\bf U}]}$,
corresponding to the zero eigen-value.

 The question studied in this paper can in fact be formulated
as follows: do the $\epsilon$-terms of the Poisson brackets of the
functionals
$$\left( {\bf S} ({\bf X}), \, \bm{\Sigma} ({\bf X}), \,
U^{1} ({\bf X}), \dots, U^{m + n_{2} + s} ({\bf X}) \right)$$
on ${\mathcal K}$ give a Poisson structure for the Whitham system
(\ref{SSigmaUWhithSyst})? It appears that this question can be
associated actually with the procedure of the Dirac restriction of
the Poisson bracket on a submanifold
(\cite{izvestia, DNMultDim, JMPMultDim}). Thus, the positive answer to
this question depends on the resolvability of the systems
\begin{multline}
\label{giJqSigmapSyst}
{\hat B}^{ij}_{[0]} ({\bf X}) \, \beta_{j[{\bf q}]} \left(
{{\bf S} ({\bf X}) \over \epsilon} + \bm{\theta}, \, {\bf X} \right)
\,\, = \,\, \left. \left\{ g^{i} (\bm{\theta}, \, {\bf X}) \, , \,
J_{[{\bf q}]} \right\} \right|_{{\mathcal K} [1]} \,\,\, ,   \\
{\hat B}^{ij}_{[0]} ({\bf X}) \,\, \alpha_{j[{\bf p}]} \left(
{{\bf S} ({\bf X}) \over \epsilon} + \bm{\theta}, \, {\bf X} \right)
\,\, = \,\, \left. \left\{ g^{i} (\bm{\theta}, \, {\bf X}) \, , \,
\Sigma_{[{\bf p}]} \right\} \right|_{{\mathcal K} [1]}
\,\,\, , \,\,\,\,\,\,\,\,\,\,\,\,\,\,\,
\end{multline}
at every ${\bf X}$, where
\begin{multline*}
{\hat B}^{ij}_{[0]} ({\bf X}) \,\, = \,\, \sum_{l_{1}, \dots, l_{d}}
B^{ij}_{(l_{1} \dots l_{d})} \left(
{{\bf S} ({\bf X}) \over \epsilon} + \bm{\theta}, \, {\bf X} \right)
\,\, \times  \\
\times \,\, k^{\alpha^{1}_{1}}_{1}({\bf X}) \dots
k^{\alpha^{1}_{l_{1}}}_{1}({\bf X})
\, \dots \,
k^{\alpha^{d}_{1}}_{d}({\bf X}) \dots
k^{\alpha^{d}_{l_{d}}}_{d}({\bf X}) \,\,\,
{\partial \over \partial \theta^{\alpha^{1}_{1}}} \dots
{\partial \over \partial \theta^{\alpha^{1}_{l_{1}}}} \, \dots \,
{\partial \over \partial \theta^{\alpha^{d}_{1}}} \dots
{\partial \over \partial \theta^{\alpha^{d}_{l_{d}}}}
\end{multline*}
is the Hamiltonian operator (\ref{MultDimPBr}) on the family
$\Lambda$, and the right-hand parts of systems (\ref{giJqSigmapSyst})
are given by the first non-vanishing terms of the brackets of
constraints with the functionals $J_{[{\bf q}]}$,
$\Sigma_{[{\bf p}]}$ on ${\mathcal K}$, having the order $O (\epsilon)$
according to (\ref{giJq}) - (\ref{giSigmap}).

 The resolvability of systems (\ref{giJqSigmapSyst}) obviously
depends on the properties of the operator
${\hat B}^{ij}_{[0]} ({\bf X})$ on the torus $\mathbb{T}^{m}$.
Easy to see that for generic values of
$({\bf k}_{1}({\bf X}), \dots, {\bf k}_{d}({\bf X}))$ the
foliation leaves defined by the set $\{ {\bf k}_{q} ({\bf X}) \}$
are everywhere dense in $\mathbb{T}^{m}$. However, we can see that
for special values of ${\bf k}_{q} ({\bf X})$ the closures of
orbits of the abelian group generated
by the set of constant vector fields
$({\bf k}_{1}({\bf X}), \dots, {\bf k}_{d}({\bf X}))$ on
$\mathbb{T}^{m}$ can define tori of lower dimensions
$\mathbb{T}^{k} \subset \mathbb{T}^{m}$. Let us denote here by 
${\mathcal M}$ the subset in the space of parameters on $\Lambda$ 
corresponding to the generic case $\mathbb{T}^{k} = \mathbb{T}^{m}$ for 
the corresponding values of ${\bf k}_{q}$. It is easy to see that the
subset ${\mathcal M}$ has the full measure in the space of parameters
on $\Lambda$.

 In general case, the operator ${\hat B}^{ij}_{[0]}$ has a finite 
number of ``regular'' eigen-vectors with zero eigen-values, smoothly
depending on the parameters on $\Lambda$. However, for special
values of parameters the set of linearly independent eigen-vectors
of ${\hat B}^{ij}_{[0]}$
with zero eigen-values can be infinite, which is connected,
in particular, with the dimension of the closures of foliation
leaves defined by the set $\{ {\bf k}_{q} \}$.

 Easy to see that, according to our definition of the quasiperiodic
function and Definition 2.2, the vectors
\begin{equation}
\label{KernelB}
v_{i}^{(l)} \left( \bm{\theta} + \bm{\theta}_{0}, \,
{\bf k}_{1}, \dots, {\bf k}_{d}, \,
{\bf p}_{1}, \dots, {\bf p}_{d}, \, {\bf U} \right) \,\,\, = \,\,\, 
\sum_{\nu=1}^{Q} \gamma_{\nu}^{l} \left(
{\bf k}_{1}, \dots, {\bf k}_{d}, \,
{\bf p}_{1}, \dots, {\bf p}_{d}, \, {\bf U} \right) \,\,
\left. {\delta J^{\nu} \over \delta \varphi^{i} (\bm{\theta})}
\right|_{\Lambda}
\end{equation}
$l = 1, \dots, s$, represent the regular eigen-vectors of the
operator ${\hat B}^{ij}_{[0]}$ on $\Lambda$, corresponding to the
zero eigen-value. It is not difficult to see also that on the
set ${\mathcal M}$ vectors (\ref{KernelB}) are the only linearly
independent kernel vectors of ${\hat B}^{ij}_{[0]}$ smoothly
depending on $\bm{\theta}$. Thus, we can see that vectors
(\ref{KernelB}) give in fact the full set of the linearly
independent regular kernel vectors of the operator
${\hat B}^{ij}_{[0]}$ on $\Lambda$.

 Using relations (\ref{KernelB}) we can claim that the right-hand
parts of systems (\ref{giJqSigmapSyst}) are automatically
orthogonal to the regular kernel vectors of the operator
${\hat B}^{ij}_{[0]} ({\bf X})$ on $\Lambda$ in the presence
of a complete Hamiltonian set of the first integrals
$( I^{1}, \dots, I^{Q} )$. Indeed, according to
(\ref{ConstrDepend}), the convolution of any Poisson bracket
$\{ g^{i} (\bm{\theta}, {\bf X}) \, , \, F \}|_{\mathcal K}$
with the variation derivatives
$\delta U^{\gamma} ({\bf Z}) / \delta \varphi^{i}
(\bm{\theta}, {\bf X})$, $\, \gamma = 1, \dots, m + n_{2} + s$,
on ${\mathcal K}$ are identically equal to
zero. It's not difficult to get then, that for
$F = J_{[{\bf q}]}$ or $F = \Sigma_{[{\bf p}]}$ this property
gives in the main order in $\epsilon$ the orthogonality of the
right-hand parts of (\ref{giJqSigmapSyst}) to the corresponding
variation derivatives (\ref{VarDer}) at every given ${\bf X}$.
As we mentioned already, the variation derivatives of the
corresponding functionals $J^{\gamma}$ give in fact the maximal
linearly independent subset in the space generated by
(\ref{VarDer}), so we get actually the same property for all
the vectors (\ref{VarDer}). From relations (\ref{KernelB}) we
get then the analogous property for the vectors
${\bf v}^{(l)} (\bm{\theta}, {\bf X})$.

 In particular, we can claim that systems (\ref{giJqSigmapSyst})
are always resolvable in the case of one-phase regular
Hamiltonian family $\Lambda$ with arbitrary number of
pseudo-phases. Indeed, the operators
${\hat B}^{ij}_{[0]} ({\bf X})$ represent in this case
skew-symmetric operators with regular spectra, such that the
nonzero eigen-values of ${\hat B}^{ij}_{[0]} ({\bf X})$ are
separated from zero.

 It can be noted also that in some examples the operators
${\hat B}^{ij}_{[0]} ({\bf X})$ do not contain a differential part 
and reduce to ultralocal operators acting separately at different 
points of $\mathbb{T}^{m}$. The corresponding systems 
(\ref{giJqSigmapSyst}) represent in this case pure algebraic 
systems and are usually trivially solvable.
The consideration of the multi-phase situation is not different
then from the single-phase one in the case of existence of the
multi-phase solutions.  Nevertheless, in the general case
operators ${\hat B}^{ij}_{[0]} ({\bf X})$ have more complicated
structure described above. As a result, systems
(\ref{giJqSigmapSyst}) can be not solvable on the space of
$2\pi$-periodic in $\bm{\theta}$ functions for some ``resonant''
values of the parameters on $\Lambda$.

 Let us formulate the Theorem which permits actually not to
separate the single-phase and the multi-phase cases in the
known examples.

\vspace{0.2cm}

{\bf Theorem 2.1.}

{\it
 Let system (\ref{InSyst}) be a local Hamiltonian system
generated by the functional (\ref{MultDimHamFunc}) according
to the Hamiltonian structure (\ref{MultDimPBr}).
Let $\Lambda$ be a regular Hamiltonian family of
$m$-phase solutions of system (\ref{InSyst}) with $n_{2}$
pseudo-phases and $(I^{1}, \dots , I^{Q})$
represent a complete Hamiltonian set of commuting integrals
(\ref{Inuvarphi}) for this family, invariant under the
action of the pseudo-phase group.

 Let the parameter space $\, {\bf U}$ contain
a dense set ${\mathcal S} \subset {\mathcal M}$ where the systems
(\ref{giJqSigmapSyst}) are solvable on the space of the 
smooth $2\pi$-periodic in each $\theta^{\alpha}$ functions.
Then:

1) The bracket
\begin{equation}
\label{AvPseudoPhaseBracket}
\begin{array}{c}
\left\{ S^{\alpha} ({\bf X}) \, , \, S^{\beta} ({\bf Y}) \right\}
 =  0 \,\,\,\,\, , \,\,\,\,\,\,\,\,
\left\{ \Sigma^{j} ({\bf X}) \, , \, \Sigma^{l} ({\bf Y}) \right\}
 =  0 \,\,\,\,\, , \,\,\,\,\,\,\,\,
\left\{ S^{\alpha} ({\bf X}) \, , \, \Sigma^{l} ({\bf Y}) \right\}
 =  0 \,\,\, ,   \\  \\
\left\{ S^{\alpha} ({\bf X})  \, , \,
U^{\gamma} ({\bf Y}) \right\} \, = \,
\omega^{\alpha\gamma}
\left({\bf S}_{\bf X}, \, \bm{\Sigma}_{\bf X}, \,
{\bf U} ({\bf X}) \! \right)
\, \delta ({\bf X} - {\bf Y}) \,\,\, ,  \\  \\
\left\{ \Sigma^{j} ({\bf X})  \, , \,
U^{\gamma} ({\bf Y}) \right\} \, = \,
\Omega^{j \gamma}
\left({\bf S}_{\bf X}, \, \bm{\Sigma}_{\bf X}, \,
{\bf U} ({\bf X}) \! \right)
\, \delta ({\bf X} - {\bf Y}) \,\,\, ,  \\  \\
\left\{ U^{\gamma} ({\bf X})\, , \, U^{\rho} ({\bf Y}) \right\}
\,\,\, = \,\,\, \langle A^{\gamma\rho}_{10\dots0} \rangle
\left({\bf S}_{\bf X}, \, \bm{\Sigma}_{\bf X}, \,
{\bf U} ({\bf X}) \! \right) \,\,
\delta_{X^{1}} ({\bf X} - {\bf Y}) \,\,\, + \,\, 
\dots \,\, + \\  \\
+ \,\,\, \langle A^{\gamma\rho}_{0\dots01} \rangle
\left({\bf S}_{\bf X}, \, \bm{\Sigma}_{\bf X}, \,
{\bf U} ({\bf X}) \! \right) \,\,
\delta_{X^{d}} ({\bf X} - {\bf Y}) \,\,\, +  \\  \\
+ \,\,\, \left[ \langle Q^{\gamma\rho \, p} \rangle
\left({\bf S}_{\bf X}, \, \bm{\Sigma}_{\bf X}, \,
{\bf U} ({\bf X}) \! \right)
\right]_{X^{p}} \,\,\, \delta ({\bf X} - {\bf Y})
\,\,\,\,\,\,\,\, , \,\,\,\,\,\,\,\,\,\,\,\,\,\,\,
\gamma, \rho \, = \, 1, \dots , m + s
\end{array}
\end{equation}
obtained with the aid of the functionals 
$(I^{1}, \dots , I^{Q})$, satisfies the Jacobi identity.

2) The averaged Hamiltonian structure is invariant with respect
to the choice of the functionals
$(I^{1}, \dots , I^{m + n_{2} + s})$ among the set
$(I^{1}, \dots , I^{Q})$ (and the choice of the set
$(I^{1}, \dots , I^{Q})$).
}

\vspace{0.2cm}

 The proof of Theorem 2.1 coincides in detail with the proofs of
Theorems 3.1 and 3.2 in \cite{JMPMultDim}, given in the absence
of the pseudo-phases. Let us say, that the considerations represented
in \cite{JMPMultDim} can be repeated without substantial changes
also in the presence of the pseudo-phases considered in the way
described above. So, let us omit here the proof of Theorem 2.1
and just make a reference to \cite{DNMultDim, JMPMultDim}.

 Thus, Theorem 2.1 gives us a possibility to generalize the
bracket averaging procedure to the case of the presence of the
pseudo-phases.

 Using Theorem 2.1, it is easy to prove that the Whitham system
(\ref{SSigmaUWhithSyst}) is Hamiltonian with respect to the
averaged Poisson bracket (\ref{AvPseudoPhaseBracket}) with the
Hamiltonian functional
\begin{equation}
\label{AveragedH}
H_{av} \,\,\, = \,\,\, \int \langle P_{H} \rangle
\left({\bf S}_{\bf X}, \, \bm{\Sigma}_{\bf X}, \,
{\bf U} ({\bf X}) \! \right) \,\, d^{d} X
\end{equation}

 Indeed, including the Hamiltonian $H$ in the set of the
functionals \linebreak
$(I^{1}, \dots , I^{m + n_{2} + s})$ it is easy to see that the
functional (\ref{AveragedH}) generates system
(\ref{SSigmaUWhithSyst}) according to bracket
(\ref{AvPseudoPhaseBracket}).

\vspace{0.2cm}

 Finally, we consider here just a very simple example of the
generalized nonlinear Shr\"odinger equation in $d \geq 1$
dimensions:
\begin{equation}
\label{MultDimNLS}
i \psi_{t} \,\, = \,\, \Delta \psi \,\, + \,\,
V^{\prime} \left( | \psi |^{2} \right) \psi
\end{equation}

 Equation (\ref{MultDimNLS}) is Hamiltonian with respect to
the Poisson bracket
\begin{equation}
\label{MDNLSBracket}
\left\{ \psi ({\bf x}) \, , \, \bar{\psi} ({\bf y}) \right\}
\,\, = \,\, i \delta ({\bf x} - {\bf y}) \,\,\, , \,\,\,\,\,
\left\{ \psi ({\bf x}) \, , \, \psi ({\bf y}) \right\}
\,\, = \,\, 0 \,\,\, , \,\,\,\,\, \left\{ \bar{\psi} ({\bf x})
\, , \, \bar{\psi} ({\bf y}) \right\} \,\, = \,\, 0
\end{equation}
with the local Hamiltonian functional
\begin{equation}
\label{MDNLSHamilt}
H \,\,\, = \,\,\, \int P_{H} ({\bf x}) \,\, d^{d} x
\,\,\, = \,\,\, \int \left( \nabla \psi \, \nabla \bar{\psi}
\,\, - \,\, V \left( | \psi |^{2} \right) \right) \, d^{d} x
\end{equation}

 Bracket (\ref{MDNLSBracket}) has $d$ momentum functionals
\begin{equation}
\label{MDNLSMoment}
I_{q} \,\,\, = \,\,\, \int P_{q} ({\bf x}) \,\, d^{d} x
\,\,\, = \,\,\, {i \over 2} \, \int \left(
\psi \, \bar{\psi}_{x^{q}} \,\, - \,\, \psi_{x^{q}} \,
\bar{\psi} \right) \, d^{d} x
\end{equation}
and the ``particle number'' functional
\begin{equation}
\label{MDNLSPartNum}
N \,\,\, = \,\,\, \int P_{N} ({\bf x}) \,\, d^{d} x
\,\,\, = \,\,\, \int \psi \, \bar{\psi} \,\, d^{d} x
\end{equation}
commuting with the Hamiltonian (\ref{MDNLSHamilt}) and with
each other. Bracket (\ref{MDNLSBracket}) is non-degenerate,
and in general the functionals (\ref{MDNLSHamilt}) -
(\ref{MDNLSPartNum}) represent the full set of local
integrals of system (\ref{MultDimNLS}) for any $d \geq 1$.

 Equation (\ref{MultDimNLS}) has a natural family of
``two-phase'' solutions given by the product of two periodic
functions:
\begin{equation}
\label{TwoPhaseNLS}
\psi ({\bf x}, t) \,\,\, = \,\,\,
e^{i ({\bf p} {\bf x} + \Omega t + \tau_{0})} \,\,
\Phi ({\bf k} {\bf x} + \omega t + \theta_{0})
\end{equation}
where the complex-valued function $\Phi (\theta)$
satisfies the equation
\begin{equation}
\label{NLStwophase}
\left( \Omega - {\bf p}^{2} \right) \, \Phi \,\, + \,\,
i \, (2 \, {\bf p} {\bf k} \, - \, \omega) \,\, \Phi_{\theta}
\,\, + \,\, {\bf k}^{2} \, \Phi_{\theta\theta} \,\, + \,\,
V^{\prime} \left( | \Phi |^{2} \right) \Phi \,\,\, = \,\,\, 0
\end{equation}

 Equation (\ref{NLStwophase}) is equivalent to the system
$$i \, (2 \, {\bf p} {\bf k} \, - \, \omega)
\,\, \Phi \, \bar{\Phi} \,\, + \,\,
{\bf k}^{2} \, \left( \Phi_{\theta} \, \bar{\Phi}
\, - \, \Phi \, \bar{\Phi}_{\theta} \right)
\,\,\, = \,\,\, i \, A \,\, , $$
$$\left( \Omega - {\bf p}^{2} \right) \, \Phi \, \bar{\Phi}
\,\, + \,\, {\bf k}^{2} \, \Phi_{\theta} \, \bar{\Phi}_{\theta}
\,\, + \,\, V \left( | \Phi |^{2} \right)
\,\,\, = \,\,\, B \,\, ,$$
where $A$ and $B$ are two real constants which are fixed by the
$2\pi$-periodicity conditions for the functions
${\rm Re} \, \Phi$ and ${\rm Im} \, \Phi$. After fixing of the
constant ``complex phase'' of the function $\Phi (\theta)$,
the form of the functions $\Phi (\theta)$
is parametrized by the three parameters
$({\bf k}^{2}, \, 2 \, {\bf p} {\bf k} \, - \, \omega, \,
\Omega - {\bf p}^{2})$ and is the same for all $d \geq 1$.

 The form of the two-phase solutions $\psi ({\bf x}, t)$
depends on the $2 d + 2$ parameters \linebreak
$(k_{1}, \dots, k_{d}, \, \omega, \, p_{1}, \dots, p_{d},
\, \Omega)$. It is very well known that the properties of the
corresponding solutions depend on the form of the potential
$V (| \psi |^{2})$. We will consider here just the possibility
of the averaging of bracket (\ref{MDNLSBracket}) on the full
family $\Lambda$ of these solutions.

 It is easy to see that system (\ref{MultDimNLS}) can be
represented as a system with a pseudo-phase. Indeed, putting
$\psi = \rho \, e^{i\phi}$ we can represent (\ref{MultDimNLS})
in the form:
$$\rho_{t} \,\,\, = \,\,\, 2 \, \nabla \rho \, \nabla \phi
\,\, + \,\, \rho \, \Delta \phi 
\,\,\,\,\,\,\,\, , \,\,\,\,\,\,\,\,\,\,
\phi_{t} \,\,\, = \,\,\, - \, {\Delta \rho \over \rho}
\,\, + \,\, \left( \nabla \phi \right)^{2} \,\, - \,\,
V^{\prime} (\rho^{2})  $$

 The Poisson bracket (\ref{MDNLSBracket}) and the functionals
(\ref{MDNLSHamilt}) - (\ref{MDNLSPartNum}) can then be written as:
\begin{equation}
\label{NewVarBracket}
\left\{ \rho ({\bf x}) \, , \, \phi ({\bf y}) \right\}
\, = \, - \, {1 \over 2 \rho ({\bf x})} \,\,
\delta ({\bf x} - {\bf y}) \,\,\, , \,\,\,\,\,
\left\{ \rho ({\bf x}) \, , \, \rho ({\bf y}) \right\}
\, = \, 0 \,\,\, , \,\,\,\,\,
\left\{ \phi ({\bf x}) \, , \, \phi ({\bf y}) \right\}
\, = \, 0
\end{equation}
\begin{equation}
\label{NewFunc}
H  =  \! \int \!\! \left( \left( \nabla \rho \right)^{2}
+  \rho^{2} \left( \nabla \phi \right)^{2}
-  V (\rho^{2}) \right) \, d^{d} x \,\,\, , \,\,\,\,\,
I_{q}  =  \! \int \! \rho^{2} \, \phi_{x^{q}} \,\,
d^{d} x \,\,\, , \,\,\,\,\,
N  =  \! \int \! \rho^{2} \, d^{d} x
\end{equation}

 The pseudo-phase group is acting in evident way:
$\rho ({\bf x}) \, \rightarrow \, \rho ({\bf x})$,
$\phi ({\bf x}) \, \rightarrow \, \phi ({\bf x}) + \tau_{0}$,
and it is easy to see that the bracket (\ref{NewVarBracket})
and the functionals (\ref{NewFunc}) are invariant under the
action of the pseudo-phase group.

 The corresponding solutions (\ref{TwoPhaseNLS}) can now be
represented in the form (\ref{RhoRepr}) - (\ref{PhiRepr}):
\begin{equation}
\label{rhophiNLSonephase}
\begin{array}{c}
\rho ({\bf x}, t) \,\, = \,\, R \left( k_{1} x^{1}
+ \dots + k_{d} x^{d} + \omega t +
\theta_{0}, \, {\bf U} \right) \,\,\, ,  \\  \\
\phi ({\bf x}, t) \,\, = \,\, \Psi \left( k_{1} x^{1}
+ \dots + k_{d} x^{d} + \omega t +
\theta_{0}, \, {\bf U} \right) \,\, + \,\,
p_{1} x^{1} \, + \, \dots \, + \, p_{d} x^{d}
\, + \, \Omega t \, + \, \tau_{0}
\end{array}
\end{equation}
where the functions $(R (\theta), \, \Psi (\theta))$ satisfy
the system:
\begin{equation}
\label{NLSonephase}
\begin{array}{c}
\omega \, R_{\theta} \,\, = \,\, 2 \, {\bf k}^{2} \,
R_{\theta} \, \Psi_{\theta} \,\, + \,\, 2 \, {\bf k} \,
{\bf p} \, R_{\theta} \,\, + \,\, {\bf k}^{2} \, R \,
\Psi_{\theta\theta}  \,\,\, , \\   \\
\Omega \, + \, \omega \, \Psi_{\theta} \,\, = \,\,
- \, {\bf k}^{2} \, R_{\theta\theta} / R \,\, + \,\,
{\bf k}^{2} \, \Psi_{\theta}^{2} \,\, + \,\, 2 \, {\bf k} \,
{\bf p} \, \Psi_{\theta} \,\, + \,\, {\bf p}^{2} \,\, - \,\,
V^{\prime} \left( R^{2} \right) 
\end{array}
\end{equation}

 So, solutions (\ref{TwoPhaseNLS}) can be considered here
as a family of one-phase solutions with one pseudo-phase.
Using relations (\ref{NLSonephase}) it is not difficult to
construct the corresponding operator
${\hat L}_{[{\bf U}, \bm{\theta}_{0}]}$ and to prove that the
family (\ref{rhophiNLSonephase}) represents a regular
Hamiltonian family of one-phase solutions with one
pseudo-phase. It is not difficult to check also that the
functionals (\ref{NewFunc}) provide a coordinate system on any
submanifold given by the constraints $p_{1} = {\rm const}$,
$\dots$, $p_{d} = {\rm const}$ in the space of parameters,
and that (\ref{NewFunc}) represent a complete Hamiltonian set
of commuting integrals according to Definition 2.2.

 Thus, we can claim that the procedure of constructing of the
averaged Poisson bracket on the family (\ref{rhophiNLSonephase})
is well justified in our case.

 For the construction of the averaged Poisson bracket we can
use just two integrals from the set (\ref{NewFunc}). All the
calculations can in fact be made in the initial coordinates
$\psi ({\bf x})$ for the bracket (\ref{MDNLSBracket}). It is
most convenient to take the integral $N$ and one of the
integrals $I_{q}$ to construct the multi-dimensional averaged
bracket. For the Poisson brackets of the densities
$P_{N} ({\bf x})$, $P_{q} ({\bf x})$ we get the following
relations:
$$\left\{ P_{N} ({\bf x}) ,  P_{N} ({\bf y}) \right\}
\, = \, 0 \,\, , \,\,\,
\left\{ P_{N} ({\bf x})  ,  P_{q} ({\bf y}) \right\}
\,\, = \,\, P_{N} ({\bf x}) \, \delta_{x^{q}}
({\bf x} - {\bf y}) \,\, + \,\, P_{N, x^{q}} \,
\delta ({\bf x} - {\bf y}) \,\, , $$
$$\left\{ P_{q} ({\bf x})  ,  P_{q} ({\bf y}) \right\}
\,\, = \,\, 2 \, P_{q} ({\bf x}) \, \delta_{x^{q}}
({\bf x} - {\bf y}) \,\, + \,\, P_{q, x^{q}} \,
\delta ({\bf x} - {\bf y}) $$

 It is easy to get also the relations $\omega_{N} = 0$,
$\Omega_{N} = 1$, $\omega_{q} = k_{q} = S_{X^{q}}$,
$\Omega_{q} = p_{q} = \Sigma_{X^{q}}$ for the frequencies
corresponding to the flows generated by the functionals
$N$ and $I_{q}$ on the family $\Lambda$.

 As a result, we define the averaged Poisson bracket on the
space of fields \linebreak
$( S ({\bf X}), \, \Sigma ({\bf X}), \, U^{1} ({\bf X}), \,
U^{2} ({\bf X}))$:
\begin{equation}  
\label{qAvBracket}
\begin{array}{c}
\left\{ S ({\bf X}) \, , \, S ({\bf Y}) \right\} \,\, = \,\,
\left\{ \Sigma ({\bf X}) \, , \, \Sigma ({\bf Y}) \right\}
\,\, = \,\, \left\{ S ({\bf X}) \, , \,
\Sigma ({\bf Y}) \right\} \,\, = \,\, 0 \,\,\, ,  \\  \\
\left\{ S ({\bf X}) \, , \, U^{1} ({\bf Y}) \right\}
\,\, = \,\, 0 \,\,\, , \,\,\,\,\,
\left\{ \Sigma ({\bf X}) \, , \, U^{1} ({\bf Y}) \right\}
\,\, = \,\, \delta ({\bf X} - {\bf Y}) \,\,\, ,  \\  \\
\left\{ S ({\bf X}) \, , \, U^{2} ({\bf Y}) \right\}
\,\, = \,\,  S_{X^{q}} \, \delta ({\bf X} - {\bf Y})
\,\,\, , \,\,\,\,\,
\left\{ \Sigma ({\bf X}) \, , \, U^{2} ({\bf Y}) \right\}
\,\, = \,\, \Sigma_{X^{q}} \,
\delta ({\bf X} - {\bf Y}) \,\,\, ,  \\  \\
\left\{ U^{1} ({\bf X}) \, , \, U^{1} ({\bf Y}) \right\}
\,\, = \,\, 0 \,\,\, , \\  \\
\left\{ U^{1} ({\bf X}) \, , \, U^{2} ({\bf Y}) \right\}
\,\,\, = \,\,\, U^{1} ({\bf X}) \,\,
\delta_{X^{q}} ({\bf X} - {\bf Y}) \,\,\, + \,\,\,
U^{1}_{X^{q}} \,\, \delta ({\bf X} - {\bf Y}) \,\,\, , \\  \\
\left\{ U^{2} ({\bf X}) \, , \, U^{2} ({\bf Y}) \right\}
\,\,\, = \,\,\, 2 \, U^{2} ({\bf X}) \,\,
\delta_{X^{q}} ({\bf X} - {\bf Y}) \,\,\, + \,\,\,
U^{2}_{X^{q}} \,\, \delta ({\bf X} - {\bf Y}) \,\,\, , 
\end{array}
\end{equation}
where $U^{1} \equiv \langle P_{N} \rangle$,
$\, U^{2} \equiv \langle P_{q} \rangle$.

 It is not difficult to check by direct calculation that after
the introduction of the action variables
$$Q_{1} ({\bf X}) \,\, = \,\, \left. \left( U^{2} ({\bf X})
\,\, - \,\, \Sigma_{X^{q}} \,\, U^{1} ({\bf X}) \right) \,
\right/ S_{X^{q}} \,\,\,\,\, , \,\,\,\,\,\,\,\, Q_{2} ({\bf X})
\,\, = \,\, U^{1} ({\bf X}) $$
the bracket (\ref{qAvBracket}) acquires the canonical form.
It is easy to check also that the action variables represent in
fact the same functionals for all the brackets (\ref{qAvBracket})
with different $q = 1, \dots, d$. Thus, all the brackets
(\ref{qAvBracket}) represent in fact the same bracket in
different coordinates. It can be also checked that the averaging
procedure gives also the same bracket for any other choice of
the pair of functionals from the set (\ref{NewFunc}). The
Whitham system is generated by the Hamiltonian functional
$$H_{av} \,\,\, = \,\,\, \int \langle P_{H} \rangle \,\, d^{d} X
\,\,\, \equiv \,\,\, \int \langle P_{H} \rangle \left(
S_{\bf X}, \, \Sigma_{\bf X}, \, U^{1} ({\bf X}), \,
U^{2} ({\bf X}) \right) \, d^{d} X $$

\vspace{0.3cm}

 The author expresses his deep gratitude to Prof. S.P. Novikov,
who introduced him to the subject of the Hamiltonian properties
of equations of slow modulations and set the problem of
justification of the bracket averaging procedure.

\vspace{0.2cm}

 The work was financially supported by the Russian Federation 
Government Grant No. 2010-220-01-077, Grant of the President
of Russian Federation NSh-4995.2012.1, and Grant RFBR No.
13-01-12469-ofi-m-2013.

\end{document}